\documentclass[11pt]{article}
\usepackage{amssymb,amsmath}
\usepackage{bm} 
\usepackage{booktabs} 
\usepackage{array}
\usepackage{latexsym}
\usepackage{graphicx}
\usepackage{color}
\usepackage{datetime}
\usepackage[nosort]{cite}
\usepackage{verbatim}
\usepackage{enumerate}
\usepackage{chngpage} 
\usepackage{mathrsfs}
\usepackage{euscript}
\usepackage{psfrag}

\usepackage{mciteplus}

\usepackage[colorlinks=true,      linkcolor=blue,      urlcolor=blue,
            filecolor=blue,      citecolor=blue,       pdfstartview=FitH,
                        pdfpagemode=UseNone,      bookmarksopen=true]{hyperref}  
\usepackage[all]{hypcap}     

\usepackage{tensor}
\usepackage{physics}
\usepackage{tikz}

\renewcommand{\comm}[1]{} 

\def\({\left(}
\def\){\right)}
\def\[{\left[}
\def\]{\right]}

\def\eg{{e.g.}}

\def\eff{{\rm eff}}

\def\coeff#1#2{{\textstyle \frac{#1}{#2}}}

\def\One{{\hbox{ 1\kern-.8mm l}}}

\def\barray{\begin{array}}
\def\earray{\end{array}}
\def\be{\begin{equation}}
\def\ee{\end{equation}}
\def\bea{\begin{eqnarray}}
\def\eea{\end{eqnarray}}
\def\bal{\begin{align}}
\def\eal{\end{align}}

\def\nBPS#1{$\frac{1}{#1}$-BPS}

\numberwithin{equation}{section} 

\definecolor{cardinal}{rgb}{0.6,0,0}
\definecolor{darkgreen}{rgb}{0,0.4,0}
\definecolor{golden}{rgb}{0.92, 0.7, 0}
\definecolor{midnight}{rgb}{0, 0, 0.5}
\definecolor{darkblue}{rgb}{0, 0, 0.7}
\definecolor{darkred}{rgb}{0.6, 0, 0}
\definecolor{purple}{rgb}{0.5, 0, 0.5}

\def\IR{\mathbb{R}}
\def\IS{\mathbb{S}}
\def\IT{\mathbb{T}}
\def\IZ{\mathbb{Z}}

\def\cE{{\cal E}}

\def\cN{{\cal N}}

\def\nBPS#1{$\frac{1}{#1}$-BPS}

\def\nicebox#1{\bigskip \framebox{\parbox{5.8 in}{{\it #1}}}\bigskip }

\begin{document}

\phantom{AAA}
\vspace{-15mm}

%
%

\vspace{1.9cm}

\begin{center}

{\huge {\bf Fuzzballs and Microstate Geometries:}}\\[.5cm]
{\huge {\bf Black-Hole Structure in String Theory }}

\vspace{1cm}

{\large{\bf {Iosif Bena$^1$,~Emil J. Martinec$^{2}$ ,~Samir D. Mathur$^{3}$  and  Nicholas P. Warner$^{1,4,5}$}}}

\vspace{1cm}

$^1$Universit\'e Paris Saclay, CNRS, CEA,\\
Institut de Physique Th\'eorique,\\
91191, Gif-sur-Yvette, France \\[8 pt]

$^2$Enrico Fermi Institute\ and Department\ of Physics, \\
University of Chicago,  5640 S. Ellis Ave.,
Chicago, IL 60637-1433, USA\\[8 pt]

$^3$Department of Physics,\\ 
The Ohio State University,\  Columbus,  OH 43210, USA \\[8 pt]

$^4$Department of Physics and Astronomy, 
and $^5$Department of Mathematics \\
University of Southern California,  Los Angeles, CA 90089-0484, USA

\vspace{0.6cm} 
{
\small
\upshape\ttfamily 
iosif.bena @ ipht.fr~,~
e-martinec @ uchicago.edu~,\\
mathur.16 @ osu.edu~,~
warner @ usc.edu} \\

\vspace{1.5cm}
 
{\textsc{Abstract}}\\
\end{center}
The black-hole information paradox provides one of the sharpest foci for the conflict between quantum mechanics and general relativity and has become the proving-ground of would-be theories of quantum gravity.   String theory has made  significant progress in resolving this paradox, and has led to the fuzzball and microstate geometry programs.  The core principle of these programs is that horizons and singularities only arise if one tries to describe gravity using a theory that has too few degrees of freedom to resolve the physics.  String theory has sufficiently many degrees of freedom and this naturally leads to fuzzballs and microstate geometries: The reformation of black holes into objects with neither horizons nor singularities. This
not only resolves the paradox but provides new insights into the microstructure of black holes.  We summarize the current status of this approach and describe future prospects and additional insights that are now within reach.  This paper is an expanded version of our  Snowmass White Paper \cite{Bena:2022ldq}.


\begin{adjustwidth}{3mm}{3mm} 
 
\vspace{-1.4mm}
\noindent
%
%

%
\end{adjustwidth}

\thispagestyle{empty}
\newpage


\baselineskip=17pt
\parskip=5pt

\setcounter{tocdepth}{2}
\tableofcontents

\baselineskip=15pt
\parskip=3pt

\newpage

\section{Overview}
\label{sec:Overview}

LIGO has now detected over 50 mergers of black holes, and the Event Horizon Telescope (EHT) is producing images of accretion disks with higher and higher resolution.   These observations are utterly remarkable as direct measurements of black holes, but, for a theorist, they also represent a phenomenal validation of General Relativity (GR).  The wave-forms measured by LIGO show that GR  captures the highly non-linear extremes of black-hole mergers with stunning fidelity.  It is therefore very natural to ask, why, at such a pinnacle of success, should we be seeking to replace GR with a ``better'' theory of gravity.

We have been here before: 110 years ago.  

Rutherford's experiment revealed the detailed internal structure of the atom, and so  a complete understanding of the  structure of matter seemed to be at hand.  There were just a few, seemingly small,  discrepant details having to do with atomic spectra.  The understanding of these ``details'' would lead to the invention of quantum mechanics and quantum field theory.   

 The frustrating, and deeply challenging, details of black hole physics, and their description in GR, are encapsulated in Hawking's information paradox.  Resolving these ``details" is going lead to a radical revision of our understanding of black-hole event horizons, change the way we think about the fundamental structure of matter and perhaps rewrite our understanding of the structure of space and time. 

Black holes are defined to be objects whose gravity is strong enough to trap light. In GR, a black hole not only traps light but also traps all other matter and information inside a surface of no return: the event horizon. As a result, the exterior of a black hole (outside the horizon) is independent of how, and from what, the black hole formed.  Moreover, the exterior structure of a black hole is completely determined by its long-range parameters, like mass, charge and angular momentum.  This is black hole uniqueness.   This is also directly related to the fact that, in GR, 
 the horizon region is a vacuum: anything near the horizon will be swept into the center of the black-hole in the time it takes light to cross the scale of the horizon ($\sim 10^{-5}$ seconds for a solar mass black hole).

When first formulated in GR, black holes were thought to be unphysical artifacts of imposing symmetry on solutions to Einstein's equations. However, in the early 1960's, Penrose's singularity theorem began the shift towards today's paradigm: black holes are essential parts of Nature.  Indeed, by the 1980's, black holes provided the best description of certain `exotic' binary star systems as well as the engines powering jets from the cores of active galaxies. The LIGO detection of black-hole mergers was thus the crowning achievement of over a century of theoretical development, finally confirmed by observations of characteristic, extremely strong-field signatures of the black holes of GR.

The frustrating details emerge when black-hole physics is combined with quantum mechanics.  In 1975, Hawking showed that the correct description of the vacuum around an event horizon leads to the emission of Hawking Radiation as a form of vacuum polarization. Because this radiation originates from just above the horizon, the uniqueness of black holes in GR implies that Hawking radiation is universal, thermal and (almost) featureless. In particular, it is independent of how the black hole formed. Semi-classical back-reaction of this Hawking radiation also implies that the black hole will evaporate, albeit extremely slowly. 

This leads to the Information Paradox:  It is impossible to reconstruct the interior state of a black hole (apart from mass, charge and angular momentum) from the exterior data, and thus from the final state of the Hawking radiation.  The evaporation process cannot, therefore, be represented through a unitary transformation of states in a Hilbert space.  Hence black-hole evaporation, as predicted by GR and quantum field theory, is inconsistent with the foundational postulate of unitarity in quantum mechanics.  Based on its horizon area, the black hole at the core of the Milky Way should have about $e^{10^{90}}$ microstates.  From the outside, black-hole uniqueness implies that its state is unique, as would be the state of its Hawking radiation when the black hole evaporates. The problem is therefore vast: $e^{10^{90}} \ne 1$

In the 47 years since Hawking's original paper, there have been many attempts at resolving the information problem.  Working purely within quantum mechanics and GR has proven  to be a dead end, except to reformulate the problem and highlight its extreme difficulty.  The paradox is also not the result of some unexpected failure of semi-classical physics.  Quite the opposite: the problem is at its sharpest for large black holes.   The horizon size is proportional to the mass, $m$, of the black hole, while gravitational tidal effects at the horizon fall off as $m^{-2}$. The conventional wisdom therefore suggests that space-time in the horizon region of a large, astrophysical black hole behaves like any other (nearly) flat piece of space-time.  Moreover, any matter near the horizon will be swept inside the black hole within the light-crossing time ($\sim m$) of the black hole.  Thus, very shortly after the black hole forms, the horizon region will be a vacuum.  There are many theorems, like black-hole uniqueness, that show that it is ``impossible" to support any matter, or other structure, near the horizon of a black hole.  Therefore, within the framework of GR and quantum mechanics, the near-horizon region will behave just like a normal piece of (nearly) flat space-time quantum vacuum. However, it is precisely this vacuum and the normalcy of the space-time at the horizon that leads to Hawking radiation and thus to the information problem.

In 2009, 
one of the authors of this review used quantum information theory to show that Hawking's information paradox could not be solved incrementally: It will require radical modification of the conventional wisdom \cite{Mathur:2009hf}. Prior to this, it had been a common belief that, because the evaporation time of a black hole is staggeringly large, the information about how the black hole formed could leak out very slowly via tiny corrections to GR (or to quantum field theory). The ``small corrections theorem" \cite{Mathur:2009hf}  based on widely-accepted ideas of how quantum field theory is formulated near black holes, showed that this idea is untenable:  resolving the paradox requires ``order $1$'' changes to horizon-scale physics.  Radical reinvention has now become an imperative.

The Fuzzball and Microstate Geometry programs are based in string theory, and currently represent the only framework in which the information paradox can be resolved while including complete, general-relativistic,  quantum-gravitational effects on the microstructure.    Conversely, these frameworks suggest that black holes, and the extreme blue-shifts of probes,  will provide  the most effective way to test predictions from string theory.

In Section \ref{sec:Info} we describe the information paradox and the essential elements of the ``small corrections theorem.''  Section \ref{Sect:perspective} contains an overview of the ideas underlying fuzzballs and microstate geometries and Section  \ref{Sect:CFT} contains a broad overview of the CFT dual of a large class of microstate geometries. The current status and recent results in microstate geometries is described in  Section \ref{Sect:MGstatus}, while Section  \ref{Sect:WS} gives an overview of the results coming from exactly solvable world-sheet dynamics in some of the stringy backgrounds central to microstate geometries.   In Section   \ref{Sect:Behavior} we describe the essential differences between microstate geometries and black holes, and discuss how, despite being horizonless, microstate geometries can still exhibit scrambling and trapping of infalling matter, and how such geometries can produce an analog of Hawking radiation.  
In lieu of a conclusion, we finish with Section  \ref{Sect:future}, describing many of the potentially very fruitful avenues for future research in fuzzballs and microstate geometries. 
In Appendix \ref{App:A} we compare and contrast  the fuzzball program with some of the more recent popular approaches to the information problem, and in Appendix \ref{App:B} we discuss an older confusion, the firewall, and how its only incarnation that makes any sense is some form of hot fuzzball.  This paper is an expanded version of our Snowmass White Paper \cite{Bena:2022ldq}.

\section{The information paradox}
\label{sec:Info}

The black-hole information paradox is a very robust puzzle because its origin is very simple: the paradox arises because  gravity is attractive.  Consider a mass $M$. Place a test mass $m$ a distance $r$ away. By itself the test mass has an intrinsic energy $mc^2$. Placing it near $M$ gives a negative potential energy $-{GMm/r}$, so we may schematically write the net energy of $m$ as
\be
E\sim mc^2-{GMm\over r}
\ee
We see that for $r<GM/c^2$, the net energy of $m$ is {\it negative}. Doing this properly, using general relativity, merely changes the result by a factor of $2$:  If
\be
r<{2GM\over c^2}\equiv R_s
\ee
then the energy contributed by $m$ (as measured from infinity) is negative for a range of 4-velocities of $m$. The radius $R_s$ is called the `Schwarzschild radius,' or `horizon radius'.  

Hawking \cite{Hawking:1975vcx}  realized that this situation leads to an instability  of the quantum vacuum. The vacuum always contains particle anti-particle fluctuations. For example, an electron-positron pair with energy $\Delta E$ can emerge from the vacuum, and then disappear  after a time $\Delta t$, where $\Delta E\Delta t\lesssim \hbar$. But now consider these fluctuations around the horizon of a black hole. One member of the pair can be inside the horizon with negative energy, and the other outside with positive energy, so that the total energy of the pair is $\Delta E=0$. The uncertainty relation then yields $\Delta t  \lesssim  \infty$; which means that the pair does not have to re-annihilate since it is an on-shell fluctuation. The outer member of the pair drifts off to infinity as `Hawking radiation', while the inner member, carrying negative energy, lowers the mass of the black hole by a corresponding amount. Thus the black hole slowly evaporates away, with the radiation described by a temperature $T=(4\pi R_s)^{-1}=(8\pi GM)^{-1}$.

At first, the existence of Hawking  radiation  seemed very satisfactory. In 1973
 Bekenstein had argued  that the black hole must have an entropy $S_{Bek}\sim A/G$, where $A$ is the surface area of the black hole \cite{Bekenstein:1973ur}. The thermodynamical relation $TdS=dE$ then demands that the black hole should have a  temperature $T\ne 0$. With this, the principle of detailed balance  requires that if black holes absorb (which they do), then  they must also radiate. Classically, nothing can emerge from the black hole (the temperature is proportional to $\hbar$ when we restore units), but Hawking's process yields emission at the quantum level. Thermodynamic consistency is obtained with:
\be
S_{Bek}={A\over 4G}
\label{SBek}
\ee
(which is inversely proportional to $\hbar$).
But soon Hawking realized that the {\it process} by which the radiation was produced leads to a serious contradiction \cite{Hawking:1976ra,Hawking:1982dj}. The two members of the created pair are in an entangled state.  This entanglement can come from the charges of the quanta, or their spins, or more generally from just the occupation numbers of the field modes which describe these quanta. The precise value of the entanglement is not relevant, so we can just model the entangled state as
\be
|\psi\rangle_{pair}={1\over \sqrt{2}}\Big (|0\rangle_b|0\rangle_c+|1\rangle_b|1\rangle_c\Big ) \ .
\label{pair}
\ee
where $b$ is the outer member of the pair and $c$ the inner member. Each emission step increases the entanglement $S_{ent}$ of the radiation with the remaining black hole by $\log 2$. This leads to a sharp puzzle near the endpoint of evaporation:

\begin{enumerate}[(i)]
\item 
Suppose we assume (as Hawking did in 1975) that the black hole evaporates away completely. Then the radiation is entangled, but there is nothing that it is entangled {\it with}. Thus the entire system cannot be described by a wavefunction, but only be a density matrix; this is a violation of quantum unitarity. 
\item
We can try to avoid the conclusion in (i)  by assuming that some unknown feature of quantum gravity stops the evaporation when the black hole reaches the Planck mass, resulting in a Planck-size remnant. But the number of different internal states for this remnant is unbounded, since we could have started with a black hole of arbitrarily large mass $M$, resulting in an arbitrarily large entanglement with the remnant. We could make the problem even worse by feeding more matter into the black hole as it evaporates.  There are serious difficulties with this situation; for example all quantum loop amplitudes tend to diverge when there are infinitely many particles in the theory below a given mass. 
\end{enumerate}
This difficulty with the evaporation process is called the black hole information paradox. 

No progress on the paradox was made for decades, since each step in the argument looked robust. The `no-hair' results indicated that the black hole was described only by its conserved charges; and the region around the horizon was in the vacuum state. The pair-creation process in this vacuum uses only low-energy physics, involving horizon scale wavelengths $\lambda\sim R_s\gg \ell_p$. Thus while we are dealing with quantum theory in a non-trivial gravitational background, the computation of Hawking radiation itself seems to require no details of quantum gravity.

Given the seriousness of this problem, one might wonder why string theorists did not devote an even larger fraction of their effort in earlier years to the paradox. A major reason was a widespread belief that the difficulty could be resolved through the effect of {\it small corrections}. Hawking did a leading-order computation using quantum fields on curved space, and it is certainly possible that small (hitherto unknown)  quantum gravity effects yield a small correction to the state of the created pair
\be
|\psi\rangle_{pair\, k}={1\over \sqrt{2}}\Big (|0\rangle_{b_k}|0\rangle_{c_k}+|1\rangle_{b_k}|1\rangle_{c_k}\Big ) +|\delta \psi_k\rangle, ~~~~~\big | |\delta \psi_k\rangle \big|<\epsilon\,,
\label{pair2}
\ee
with $\epsilon\ll 1$. But the number of quanta produced by the black hole is very large
\be
N_{pairs}\sim S_{Bek} \sim \left ( {M\over m_p}\right )^2
\ee
Perhaps these small corrections in (\ref{pair2}) introduce subtle correlations among the large number of radiated quanta $b_k$, so that by the end of the evaporation the radiation state is not really entangled with the remaining black hole? If so, there would be no paradox: Hawking's leading order entanglement $S_{ent}=N_{pairs} \log 2$ would be removed by the large number of subleading correction terms.

However, in \cite{Mathur:2009hf} 
it was shown that such small corrections (\ref{pair2})  cannot resolve the problem. Suppose we assume that  
\begin{enumerate}[(i)]
\item
The correction to each pair is small as in (\ref{pair2}).
\item
Once a radiation quantum, $b_k$, has receded sufficiently far from the black hole, any further changes to the black holes' state are not relevant to the  radiation quantum.
\item
The vacuum is unique.
\end{enumerate}
Then the entanglement $S_{ent}(N)$ at step $N$ will keep rising as
\be
S_{ent}(N+1)>S_{ent}(N)+\log 2-2\epsilon
\ee
 This `small corrections theorem'  promotes Hawking's argument into a rigorous result:  small modifications to semi-classical dynamics can never resolve the paradox; we need  order-{\it unity} modifications. The reason that this result was not intuitively obvious was that its proof involves use of the strong subadditivity inequality  of quantum entanglement entropy, and this inequality does not have any elementary proof.
 
 As we will see below, when we try to make black holes in string theory, we do not obtain the semi-classical black hole. Instead we get a large object with no horizon, termed a {\it fuzzball}. Such an object radiates from its surface like a normal body, not by the process of creating pairs (\ref{pair2}) from the vacuum; thus there is no information paradox. The remarkable fact here is that quantum gravity effects have modified the semi-classical structure not within Planck distance of the singularity, but throughout  the interior of the horizon. It remains true, of course, that the natural length scale made from the fundamental constants $c, \hbar, G$ is the Planck scale $\ell_p$. But a large black hole is made of a large number of quanta $N$, and we find that the relevant scale, $L$, for quantum gravity grows as $L\sim N^\alpha \ell_p$ with $\alpha$ such that $L\sim R_S$. This remarkable fact is likely to have implications far beyond black holes; for instance it suggests the possibility of new dynamics at the scale of the cosmological horizon, thus impacting the cosmological-constant problem.

\section{Fuzzballs and microstate geometries: A broader perspective}
\label{Sect:perspective}

\subsection{Some background}
\label{ss:background}

For the first fifteen years after Hawking's presentation of the paradox, many creative new ideas were thrown at the problem, but there was no significant progress.  

The first real advance came with the work of Sen \cite{Sen:1995in} and Strominger and Vafa \cite{Strominger:1996sh}.   They showed that a string theoretic description of a black hole has the capacity to count its microstates.  This was demonstrated in a very specialized limit: The black holes were supersymmetric, and thus had vanishing Hawking temperature, and the stringy counting was done in a limit in which the string coupling, $g_s$, and hence Newton's constant, $G_N \sim g_s^2$, vanishes.  The triumph was to show that the
states of the string system have an entropy that exactly matches the horizon area of the black hole at finite $G_N$.

Vanishing Hawking temperature enables one to  focus on a crucial first step: the ``information storage problem''.  Supersymmetry means that one can count  states using index methods.  The vanishing Newton's constant seems radical, but the idea is that index states are protected by supersymmetry under smooth deformations.  Therefore the index states will be preserved as one turns on $G_N$ and so, while one does not immediately know what the microstructure becomes in the black hole, one knows that it must still be there, somewhere.  

However, to resolve the information paradox, it is not sufficient to count the microstates: One must determine their {\it structure}  and understand it at strong effective coupling, where the stringy bound state is expected to give a black hole.    The challenge here is that gravitational attraction is universal and gets stronger with $G_N$, and so matter gets ever-more compressed as $G_N$ increases.  On the other hand, the area of a black-hole horizon grows with $G_N$.  Thus, one would expect that the index states would collapse behind a horizon when gravitational compressive forces exceed the ability of matter to sustain the pressure necessary to prevent core collapse. 

String theory evades this seemingly inevitable outcome through {\it fractionation} of its spectrum on extended objects \cite{Das:1996ug,Maldacena:1996ds}.  

String theory necessarily contains solitonic branes.  When a brane, or string, is bound to a large number, $N$, of another set of branes, then the tension of the first brane becomes a fraction, $1/N$, of its normal tension.  In  \cite{Mathur:1997wb}, an estimate was made of the size of the brane bound states of Strominger and Vafa.  Remarkably, it was found that the size, $D$, of this bound state increases with the number of branes in the state, as well as with the coupling, so that we always have $D\sim R_s$, where $R_s$ is the radius of the horizon. 
Similar results were found~\cite{Banks:1997hz,Horowitz:1997fr,Banks:1997tn} in the BFSS matrix model~\cite{Banks:1996vh}, a precursor to (and in hindsight, an example of) gauge/gravity duality.
These results suggested that, in string theory, one never gets the traditional black hole with a vacuum horizon; instead one gets a horizon-sized `fuzzball', which radiates from its surface like a normal body, thus evading the information paradox.  Indeed, it was subsequently shown  \cite{Bena:2004wt} that the gravitational back-reaction implies that the types of stringy configurations counted in \cite{Strominger:1996sh} grow with $g_s$ at exactly the same rate as the horizon.  Thus, at finite $g_s$, stringy microstructure does not necessarily form horizons.

Shortly after this came the AdS/CFT correspondence and holographic field theory, which also suggested that black-hole formation and evaporation must be a unitary process.  Essentially, the formation of a black hole (at least in AdS) is dual to some evolution of states in a holographic theory on the boundary of the space-time.  Since the dual field theory is unitary, the gravitational process must also be unitary.  

Thus, given the correspondence, one has a wide variety of examples of unitary quantum dynamics of black holes.  As such, it provides an {\it existence} proof of counterexamples to Hawking's assertion that black holes violate quantum mechanical unitarity.  However, this result is far from complete: what one really needs is a {\it constructive} proof, one that tells us {\it how the bulk gravitational dynamics is modified such that it becomes unitary}.  As one solves this problem, one can also try to match the unitary evolution in the bulk with that of  the boundary, and thus develop the holographic dictionary.

Fuzzballs and microstate geometries fit into the idea of AdS/CFT duality in an obvious and natural way. The CFT has states that we can identify as states of a `black hole' phase. The gravity dual of any such microstates must be a solution of string theory with no horizon.  The absence of horizon  is essential for any individual microstate: horizon area is related to entropy in black hole thermodynamics, and an individual microstate has no entropy.

This picture was also strongly supported by developments in holographic field theory. Black holes, with their horizons and temperature, provide the gravity duals of thermal field theory and ensemble averaging. On the other hand, individual phases or states, when sufficiently coherent, are described by smooth geometries.  For example, the confining phase of a theory that is in the same universality class as QCD is dual to the smooth Klebanov-Strassler geometry \cite{Klebanov:2000hb}. More generally, one of the core ideas of string compactification re-surfaced in holography: the link between phase changes in the field theory and geometric transitions.  A singular brane source, representing one possible phase, can ``dissolve'' into  fluxes that thread new topological cycles that emerge in the core of the transitioned geometry. The fluxes on these cycles are holographically dual to the order parameters of  the emergent infra-red phase of the field theory, and the scale of the cycles provides the  scale of the new phase \cite{Bena:2004jw, Lin:2004nb, Dymarsky:2005xt}.  Thus gaugino condensates, and chiral symmetry breaking, are described by precisely such a transition from a singular brane-sourced geometry to a smooth, topologically non-trivial background.

Microstate geometries have their roots in these holographic insights.  The core idea is that, whatever the the black-hole microstates become at finite $G_N$, ground states and sufficiently coherent excitations are often describable by smooth solutions to low energy limits of string theory: supergravity.  

\subsection{Fuzzballs and microstate geometries}
\label{ss:fuzzbackground}

In our discussion we  take pains to distinguish the following two notions:
\begin{itemize}
\item
{\it Microstate geometries} are smooth horizonless solutions of the supergravity theory that is the effective long-wavelength approximation to string theory.  These solutions can have the same mass, charge and angular momentum as a black hole.
\item
{\it Fuzzballs} are the general class of objects in which the black-hole interior is supplanted by some new, quantum, horizon-scale structure having itself no horizon.  This structure is built out of the branes and other stringy ingredients that are known to account for black hole entropy at weak string coupling. 
\end{itemize}
In particular, microstate geometries are examples of fuzzballs, but not all fuzzballs need be microstate geometries.  The ``fuzz'' of fuzzballs might for instance include string or brane condensates whose wave-function extends over the horizon scale, and whose degrees of freedom are the dominant components of black-hole entropy.

It is now understood that black holes are highly chaotic systems~\cite{Shenker:2013pqa,Maldacena:2015waa}, and this is consistent with idea of a fuzzball as a horizonless, complex, stringy quantum state.  However,  
such chaos is in tension with the idea of microstate geometries, which are smooth coherent limits of fuzzballs.   One possibility is that microstate geometries  only capture states of black holes at, or near, zero temperature.  One can then try to use these ``low-temperature states'' as a basis for more general classes of black holes.  More recent calculations, like those in \cite{Ganchev:2021pgs, Ganchev:2021ewa}, also suggest that highly coherent black-hole microstates at finite temperature can be constructed by putting the system in a ``box'' created by anti-de-Sitter (AdS) boundary conditions.  Such coherent states then represent black holes in equilibrium with the radiation they produce. 

The ultimate role of microstate geometries will boil down to how generic such geometries are in the ensemble of all microstates, and whether there might be effects beyond supergravity that are crucial ingredients of horizon-scale physics.    The practical point is that generic, chaotic quantum states  are very hard to compute, while microstate geometries, and their excitations, provide an invaluable starting point for computing  properties of horizon-scale microstructure.

There are many ``no-go'' theorems that suggested that microstate geometries do not exist; however, as we will discuss in Section~\ref{Sect:MGstatus}, vast and diverse families of such geometries have now been constructed. Moreover, by dissecting the standard  ``no-go'' theorems for solitons, one can understand precisely how the known microstate geometries dodge so many bullets.  Indeed, such geometries only exist in at least $4+1$ dimensions, and the key lies in supporting the smooth geometries with topological fluxes that thread non-trivial p-cycles (with $p \ge 2$),  \cite{Gibbons:2013tqa,deLange:2015gca,Haas:2014spa}. We now know that this mechanism is the {\it only} way to create non-trivial, horizonless, time-independent solitonic solutions \cite{Gibbons:2013tqa,deLange:2015gca}.  Indeed, microstate geometries provide the only {\it gravitational} mechanism that can support horizon-scale microstructure. 

{\it Scaling} microstate geometries are the most relevant to black holes.  In general, scaling means that there is a ``scale parameter'' that controls the red-shift at which the soliton deviates significantly from the black hole and ``caps off smoothly."    For extremal microstate geometries, this scale parameter controls the depth of the AdS throat.  Such geometries lend themselves to holographic analysis, and the families of microstate geometries known as {\it superstrata} have an extremely well established holographic dictionary with the D1-D5 CFT.   Indeed, superstrata have given a precise description of a subset of the microstates of the Strominger-Vafa system at finite $G_N$.  We will return to this point in Section~\ref{Sect:CFT}.

Apart from giving one a controlled handle on microstructure,  scaling geometries can exhibit some very interesting new physics: they are well within the supergravity approximation geometrically, but the deepest of these geometries are governed, on a macroscopic scale, by quantum effects  \cite{Bena:2006kb,Bena:2007qc,deBoer:2008zn,deBoer:2009un,Martinec:2015pfa,Li:2021gbg,Li:2021utg}.

\subsection{Some general principles}
\label{ss:principles}

The primary reason for investigating fuzzballs and microstate geometries is that they provide a natural, and at least conceptually-simple, solution to  the information problem. Indeed  the {\it central precept of fuzzballs} is: 

\nicebox{Fuzzballs represent a new phase  that emerges when matter is compressed to black-hole  densities, and this new phase prevents the formation of a horizon or singularity.  A fuzzball does not have an information problem  because the internal states of the fuzzball are in causal communication with distant observers.}

While vastly more complicated, fuzzballs have no more of an information problem than do pieces of coal, white dwarfs and neutron stars.

\subsubsection{The fuzzball principle}
\label{ss:fuzzprinciple}

One of the core principles that has emerged from 20 years of study of fuzzballs and microstate geometries is the {\it ``fuzzball principle:"} 

\nicebox{Horizons and singularities only appear if one tries to describe gravity using a theory that has too few degrees of freedom to resolve the physics.}

First and most fundamental,  is the idea  that string theory  has sufficiently many degrees of freedom to resolve individual black-hole microstates into complicated, intrinsically-stringy, horizonless  objects, namely, fuzzballs.  Microstate geometries are then the coherent states within the ensemble of all black-hole, or fuzzball, microstates.   
While obviously more limited than generic fuzzballs, microstate geometries have the huge advantage that detailed computations can be done.  It is from these that we have learned much about the expected behavior of fuzzballs.    Indeed, one way to begin to explore more generic fuzzballs is to examine string excitations around microstate geometries.

There have now been several instances in which the fuzzball principle has proven extremely robust.   There are several families of microstate geometries that appear to have an instability, or a degenerate or singular corner of its moduli space, that produces a black object.   However, in each instance, it has been shown that, once one examines the instability, or apparent degeneracy/singularity, one finds that it can be resolved in terms of scrambling into stringy states, and/or through the emergence of a new branch of the moduli space of microstate geometries\footnote{Examples of this include, the trapping instability \cite{Eperon:2016cdd,Marolf:2016nwu, Bena:2020yii} that naively leads to shockwave singularities~\cite{Chakrabarty:2021sff}; the degeneration limits of supertubes~\cite{Martinec:2020gkv}; and the zero angular-momentum limit of superstrata \cite{Bena:2022sge}.  }.

\subsubsection{Fuzzball formation}
\label{ss:FuzzFormation}

One of the most natural challenges to the whole idea of fuzzballs is to set up a thought experiment whose outcome appears to be determined by some generalization of Birkhoff's theorem.  Specifically, set up a vast, perfectly-spherical shell of low-density matter, but with a titanic total mass.  How could this collapse into anything other than a Schwarzschild black hole? How  could perfect spherical symmetry not inevitably lead  to a horizon and a singularity?  More generally, General Relativity suggests that there are a vast array of initial states that must collapse to form a horizon and singularity. 

In a similar vein, we also  learn, in General Relativity, that the horizon region is in the vacuum state, and that the curvature at the horizon of a black hole of mass $m$ is proportional to $m^{-2}$.  Thus, for large black holes, the space-time in the horizon region is very close to being empty, flat space, and must therefore be completely normal.  

We believe that both of these classical intuitions are too na\"ive because they ignore the elephant in the room~-- specifically, the  huge entropy, and the concomitant vast concentration of low-energy excitations  sitting in the black hole.  

  The Bekenstein entropy (\ref{SBek}) is {\it much} larger than the entropy of normal matter that we could place within the same region with the given energy. But with the traditional picture of the black hole, it is not clear what role this large entropy could play in the dynamics of the black hole. With fuzzballs, we find that there is a non-zero probability, $P$, for the collapsing star to tunnel into a fuzzball configuration. We can estimate this probability as:
  \be
  P=|{\mathcal A}|^2, ~~~{\mathcal A}\sim e^{-S_{cl}} \,,
  \ee
   where the  the classical gravitational action is estimated by setting all length scales to be order the black hole radius $\sim GM$
  \be
  S_{cl}\sim {1\over G}\int d^4 x {\mathcal R} \sim {1\over G} (GM)^4 {1\over (GM)^2} \sim GM^2\sim \left ( {M\over m_p}\right)^2 \,.
  \ee
  Thus for a large black hole, $M\gg m_p$, the probability for tunneling, ${\mathcal P}$, is indeed very small, as expected for the tunneling between two macroscopic configurations. But we must multiply this probability by the number ${\mathcal N}$ if fuzzball states that we can tunnel to, and this number is very {\it large}
  \be
  {\mathcal N}\sim e^{S_{Bek}} \sim e^{A\over 4G}=e^{4\pi GM^2} \,.
  \ee
  We then see that the abnormally large entropy of black holes makes it is possible to have \cite{Mathur:2008kg}
  \be
  {\mathcal P}{\mathcal N}\sim 1 \,.
  \ee
  This implies that the collapsing star tunnels into a linear combination of fuzzball states instead of following the classically predicted evolution. 
  
  In \cite{Kraus:2015zda},  an indirect argument was given for the exact cancellation of exponents in  $  {\mathcal P}$ and ${\mathcal N}$. In \cite{Bena:2015dpt} an explicit example was worked out for this `entropy enhanced tunneling' within a certain family of fuzzball states.

Therefore, the answer to the issue of   fuzzball formation  is that, during collapse, a fuzzball must emerge through spreading of the quantum wave-function, governed by a quantum phase transition. The   vast density of microstates means a black hole, or fuzzball, becomes a quantum object at the horizon scale: 

\nicebox{When one compresses matter into a state in which its entropy approaches $\frac{1}{4} A$, where $A$ is its surface area in Planck units, then it becomes an intrinsically quantum object. That is, quantum effects become large at the horizon scale.}

\noindent
This is one of the most important aspects of  black-hole physics, and we will refer to it as  {\it quantum incompressibility}.  It also implies that   the space-time, at the horizon scale, is far from classical ``normality.''

\subsubsection{Collective fuzzball dynamics }
\label{ss:Collective}

The  macroscopic quantum properties of fuzzballs  raises a red flag for microstate geometries: If black holes are intrinsically quantum, then what use is classical supergravity in describing them?    There are two deeply interrelated answers to this question.

First, supergravity, because it is based on the massless sector of string theory, captures the large-scale collective effects of a fuzzball. As a result,  coherent, large-scale quantum effects should be captured by a classical correspondence limit, namely supergravity.  Moreover, one might reasonably hope that simple, semi-classical quantization of the supergravity phase space could give some further details of the quantum fuzzball.  The obvious analogies are the kinetic theory of gases and the vibrational modes of molecules.    The underlying systems are quantum, but the semi-classical analysis not only gives a good description of the bulk thermodynamics, but also gives a valuable, approximate description of the microstructure.

This basic idea has been given much deeper  support from holographic field theories:

 \nicebox{Via holography, supergravity can provide a remarkably good description of the low-energy collective effects of strongly-coupled quantum field theories.}
 
The simplest, but most celebrated example of this idea is the computation of the viscosity-to-entropy ratio in quark-gluon plasmas. (See, for example, 
\cite{Policastro:2001yc,Kovtun:2003wp,Kovtun:2004de,Heinz:2009xj}.)   However this was simply based on an effective black-hole description.  Microstate geometries are expected to provide much richer and more detailed information.  In particular,  they have many more moduli, like scales and order parameters, that control the IR physics and describe the collective excitations of the fuzzball. 
 
For microstate geometries and black holes, we will refer to this principle  as  saying that supergravity can capture the  {\it ``collective fuzzball dynamics.''}  As we will discuss in Section  \ref{Sect:MGstatus}, some of the best evidence that these ideas are correct is the success of the microstate geometry program itself, and its matching via precision holography.

Thus fuzzballs and microstate geometries 
provide a promising route to
understanding black-hole microstructure.  Moreover, even if one does not accept every aspect of the microstate geometry program,
one should note that if one wants to study horizon-scale microstructure, microstate geometries provide the only mechanism for {\it gravitationally} supporting that microstructure \cite{Gibbons:2013tqa,deLange:2015gca,Haas:2014spa}.  
It is also possible that some more stringy mechanism will emerge as one transitions from microstate geometries to intrinsically quantum fuzzballs~\cite{Martinec:2015pfa,Martinec:2019wzw,Martinec:2020gkv}.  Whatever that more general mechanism proves to be, it should emerge from the dynamics of microstate geometries.  Thus, talking about horizon-scale microstructure without microstate geometries means, quite literally, that one is  doing unsupported physics.

\section{The spacetime CFT}
\label{Sect:CFT}

Holographic duality emerges when one assembles a large number, $N$, of branes in string theory.   The low-energy, internal dynamics of the branes is described by a large-$N$ theory coupled to matter and the dynamics can have a non-trivial IR renormalization group fixed point~-- a conformal field theory.  Away from the brane, in the ``bulk,''   the brane bound state sources gravity.  The near-source limit is also an IR limit due to gravitational redshift of the near-source region.  The two descriptions~-- gravitational and ``field-theoretic''~-- yield equivalent or dual realizations of the same physics.  Conformal symmetry of the field theory is realized on the gravity side as the geometrical symmetry of AdS spacetime.  

This is a strong/weak coupling duality, which means that when the  
theory on the branes is weakly coupled, the geometry is strongly curved, and vice versa.  Gravitational thermodynamics is equivalent to the  thermodynamics of the theory on the branes, and gravitational entropy is accounted for by the underlying degrees of freedom of this theory.  For instance, in the duality between gravity in $AdS_5\times \IS^5$ and $\cN=4$ supersymmetric Yang-Mills theory (SYM), the asymptotically-$AdS_5\times \IS^5$ solution containing a black hole describes, holographically, the thermally deconfined phase of the gauge theory.  The challenge is to understand how the microstructure that is realized explicitly in the CFT, is manifested on the gravity side of the duality.

The duality underlying the counting of supersymmetric black-hole microstates in \cite{Strominger:1996sh} relates supergravity on $AdS_3\times \IS^3\times \IT^4$ to a two-dimensional CFT that is the IR limit of the open string dynamics on a bound state of $n_1$ D1-branes and $n_5$ D5-branes.  Again, when the CFT is weakly coupled, geometry is strongly curved and vice versa, and the  large number is $N=n_1n_5$.  However,  two-dimensional conformal symmetry algebra dictates~\cite{Cardy:1986ie,Strominger:1997eq} that the asymptotic density of states is determined by the central charge $c=6N$ of the conformal algebra:
\be
\label{Sbtz}
S = 2\pi\Big( \sqrt{\coeff16 c\, \cE - J^2} + \sqrt{\coeff 16 c\, \bar \cE-\bar J^2} \Big) ~,
\ee
 where $\cE,\bar \cE=E\pm P$ are the energies of left- and right-moving two-dimensional CFT excitations, and $J,\bar J$ are left- and right-moving angular momenta on $\IS^3$. 
 
This entropy can again be thought of as made up of deconfined gauge-theory degrees of freedom, but there are more degrees of freedom than either one-branes or five-branes would have on their own~-- the whole is more than the sum of its parts.  For instance, $n_1$ D1-branes carrying a gas of left-moving excitations would have an entropy $S=2\pi\sqrt{2 n_1\cE}$~\cite{Sen:1995in}; bound to the $n_5$ D5-branes, the entropy is enhanced by a factor $\sqrt{n_5}$.  This ability for various branes working together to more finely divide the available free energy into constituents ({\it mode fractionation}) provides a mechanism by which string theory can evade the compression  of gravity~-- the more fractionated the constituents are, the more spread out  their quantum wave-function becomes.

While the  D1-D5 CFT accounts for all the entropy, it doesn't tell us how that information is arranged spatially; in the regime where gravity applies, the CFT is a strongly-fluctuating, strongly-correlated theory.  Microstate geometries give examples of non-singular geometries where the holographic map between the two sides of the duality can be worked out explicitly.

The examples studied in most detail are states that preserve some amount of supersymmetry, which suppresses a variety of quantum corrections and thereby allows one to reliably compute in the strongly coupled CFT dual to weakly coupled gravity.  For instance, one can compute the partition function of supersymmetric states using the CFT at weak coupling because it receives no quantum corrections during the passage to strong coupling.

Supersymmetric (BPS) ground states of the D1-D5 system can be labelled by the data of how the D1-branes bind to the D5-branes.  The strings fractionate in the bound state, such that the total string winding charge, $n_1$, becomes fractional-string winding charge $n_1n_5$~\cite{Dijkgraaf:1997ku,Maldacena:1996ya}.  This total winding charge can be partitioned into $n_w$ fractionated strings of winding $w$; the entropy of such partitions yields the entropy of supersymmetric ground states
\be
S = 2\pi\sqrt{2 (n_1 n_5-|J|)}  ~.
\ee
Sets of states with windings $w>1$ form {\it twisted sectors} of the CFT.
Adding momentum and angular momentum excitations to these fractionated strings leads to the asymptotic density of states~\eqref{Sbtz}~\cite{Strominger:1996sh}.  

The fractionation of string winding has another important effect~-- the maximal allowed winding $w\sim n_1n_5=N$ leads to a characteristic energy gap $\delta\cE = {2 \over N R}$, where $R$ is the radius of the common circle wrapped by the D1- and D5-branes.  The excitations in sectors of large winding are highly redshifted with respect to the AdS energy scale $\delta\cE_{AdS}\sim 1/R$.  Nevertheless this tiny gap in the CFT excitation spectrum, $\delta\cE$,  regulates the near-horizon, IR physics, in the bulk.  This $1/N$ gap disappears in the classical limit $N\to\infty$~-- classical black holes have zero gap in their excitation spectrum.  We see that quantum and stringy effects are crucial for regulating the near-horizon dynamics.%
\footnote{The gap discussed here is associated to the time scale of thermalization of an infalling probe, and is distinct from the gap in the spectrum at high energies, which is of order $e^{-S}$.  States nearby in energy typically correspond to very different configurations of the effective strings, which are only reached over the exponentially long time scale needed to ergodically explore the phase space.}

The supergravity duals to these supersymmetric ground states were constructed in~\cite{%
Lunin:2001fv,
Taylor:2005db,
Kanitscheider:2007wq}.
Each ground state in the CFT is a collection of fractionated strings of various windings and polarizations, which can be mapped precisely onto a particular source configuration for the geometry.  The geometries cap off smoothly at a redshift controlled by the angular momentum on $\IS^3$ carried by the background, with each source generating a distinct field configuration.%
\footnote{Indeed, the entropy can be recovered from a geometrical quantization of supergravity phase space~\cite{Rychkov:2005ji,Krishnan:2015vha}.  Essentially, the supergravity solutions form a coherent state basis for the quantum Hilbert space.}
Heuristically, the angular momentum pries apart the source slightly, giving a tiny mass to the stringy excitations of the background branes which serves as an IR regulator.   For these ground state ``supertubes'',  the curvature in the cap of a generic source configuration is of order the string scale \cite{Lunin:2002iz, Chen:2014loa}, and so the effective supergravity theory is already starting to break down there.  {Fortunately, as we will discuss in Section~\ref{Sect:WS}, world-sheet string theory methods are available to reliably analyze the structure of the cap in ground states and their perturbations. Moreover, as we will discuss below, the supergravity approximation  is restored once one adds a significant amount of momentum charge. }

In addition to the collection of supersymmetric ground states, one can map the spectrum of BPS operators between the two sides of the duality that generate excitations~\cite{Larsen:1998xm,deBoer:1998kjm}%
\footnote{In making the comparison, one needs to carefully account for operator mixing through a holographic renormalization prescription.}.
The operators themselves mediate transitions among the ground states, leading to a description of the space of ground states as different condensates of supergravitons acting on the CFT vacuum~\cite{Lunin:2002bj}.  One often refers to such states, which differ macroscopically from the vacuum, as ``heavy'' backgrounds, and the multi-particle operators that create them from the vacuum as ``heavy'' operators.  Correspondingly, the single-particle supergravity operators are referred to as ``light'' operators.

The expectation values of CFT operators, which can be read off from the asymptotic behavior of supergravity fields, are then heavy-heavy-light (HHL) correlation functions in the CFT, which have been analyzed on both sides of the duality in~\cite{%
Jevicki:1998bm,
Mihailescu:1999cj,
Lunin:2000yv,
Lunin:2001pw,
Arutyunov:2000by,
Skenderis:2006uy,
Skenderis:2006ah,
Kanitscheider:2006zf,
Taylor:2007hs}
and exhibit perfect agreement.

To move away from the ground state, one wants to excite more general modes in the cap.  The first efforts in this direction consisted of perturbative excitations~\cite{%
Mathur:2003hj,
Giusto:2004id,
Giusto:2004ip,
Giusto:2006zi,
Giusto:2011fy,
Lunin:2012gp,
Shigemori:2013lta},
as well as coherent excitations using the current algebra symmetries of the theory~\cite{%
Jejjala:2005yu,
Mathur:2012tj,
Giusto:2012yz,
Giusto:2013rxa,
Giusto:2013bda,
Chakrabarty:2015foa}.
A major advance came with the development of tools to make a wide variety of supergravity solutions with momentum-carrying excitations, known as
{\it superstrata}~\cite{%
Bena:2015bea,
deLange:2015gca,
Bena:2016agb,
Bena:2016ypk,
Bena:2017xbt,
Bakhshaei:2018vux,
Ceplak:2018pws,
Heidmann:2019zws,
Mayerson:2020tcl,
Ganchev:2021iwy}.
Free-energy considerations indicate that the momentum preferentially populates the highly fractionated sectors. This causes the configuration to  spread dramatically, both in phase space and in spacetime. Indeed, the configuration grows with $g_s$ at the same rate as the corresponding black-hole horizon.  Superstrata are coherent expressions of these microstates and their features  and structure appear at  macroscopic scales, and are thus well-described by supergravity. 

The holographic map to the corresponding ``superstratum states'' in the CFT has been  developed and correlators in superstrata have also been computed.  In particular, HHL and HHLL correlators, which are the one- and two-point functions of light probes in the heavy supergravity background, have been thoroughly analyzed~\cite{%
Giusto:2015dfa,
Galliani:2016cai,
Bombini:2017sge,
Bombini:2019vnc,
Tian:2019ash,
Tormo:2019yus,
Giusto:2019qig,
Bena:2019azk,
Giusto:2019pxc,
Hulik:2019pwr,
Giusto:2020mup,
Ceplak:2021wzz,
Rawash:2021pik}.
Full agreement is found where a direct comparison between supergravity and CFT can be made.%
\footnote{CFT computations are typically made at weak coupling, far from the supergravity regime where the CFT is at strong coupling. Some quantities computed at weak coupling are renormalized as one moves to string coupling, which can sometimes give rise to a discrepancy between results on the two sides.}

\section{Current status of microstate geometries}
\label{Sect:MGstatus}

\subsection{Overview}
\label{ss:status-overview}

Supersymmetry, and the associated first-order BPS equations, greatly simplify the construction of microstate geometries, and so by far the most progress has been made in supersymmetric constructions.  Moreover, precision holography has focused  on supersymmetry because it protects a range of correlators that can be computed on both sides of the duality.   

Over the years, there has been some construction of non-supersymmetric microstate geometries, but these have often led to extremal, non-supersymmetric solutions, or solutions with very high angular momenta  \cite{Jejjala:2005yu, Bena:2009qv, Bobev:2011kk, Vasilakis:2011ki, Bena:2015drs,Bena:2016dbw,Bossard:2017vii,Bah:2020ogh,Bah:2020pdz,Bah:2021owp}.  However, the last year has seen some spectacular progress in the systematic construction of very interesting, non-supersymmetric microstate geometries \cite{Ganchev:2021pgs,Ganchev:2021ewa,Heidmann:2021cms,Bah:2022yji}.  These constructions demonstrate that many of the remarkable features of supersymmetric microstate geometries are shared by non-supersymmetric counterparts.   

Microstate geometries, both supersymmetric and non-supersymmetric, fall into two broad classes: (i) ``multi-centered'', bubbled geometries, with multiple non-trivial cycles whose classical microstructure is largely in the moduli space of the cycles,  and (ii) Superstrata and microstrata, in which there is  only one non-trivial cycle, and the microstructure is encoded in shape modes, or momentum waves on that cycle.   

Ultimately there should  be no distinction between these classes as we expect there to  be large families of fluctuating, multi-centered microstate geometries.  The problem is simply the technical challenge of constructing them.    Superstrata \cite{Shigemori:2013lta,Giusto:2013bda,Bena:2015bea,Bena:2016ypk,Bena:2017geu,Bena:2017upb,Bena:2017xbt,Ceplak:2018pws,Heidmann:2019xrd,Heidmann:2019zws,Walker:2019ntz,Ganchev:2021iwy} and microstrata  \cite{Ganchev:2021pgs,Ganchev:2021ewa} have received attention because they are relatively simple to analyze and probe, and their  holographic  duals are well understood.  On the other hand, multi-centered bubbling microstate geometries pose an intriguing,  unsolved problem in holography.  These geometries can be arranged to have deep, scaling AdS throats, and so must be dual to pure states in the D1-D5 CFT.  While it is conjectured that these could be  condensates of twisted sector states  \cite{Giusto:2012yz,Shigemori:2021pir}, we currently have no proof of this. Furthermore, it is possible that some of these states only preserve supersymmetry at special loci in the moduli space \cite{Bossard:2019ajg}, which further complicates their holographic interpretation.

The construction of almost all microstate geometries has focussed on five- and six-dimensional supergravities obtained by trivial compactifications of M-theory or IIB supergravity.  This means the compactification manifold is fixed, and does not participate in the dynamics, or geometric transitions, of the microstructure.  This runs somewhat counter to the index computations, which suggest that the majority of the microstructure arises from non-trivial compactification topologies.   Another of the major open problems in microstate geometries  is to put the topological transitions in the compactification manifold and the space-time on the same footing.  We believe that this is simply a technical problem.  Indeed, the recent construction of non-BPS microstate geometries  \cite{Heidmann:2021cms,Bah:2022yji}   achieves this paradigm beautifully, with topology created from cycles pinching off in both the compactification and space-time directions. 

\subsection{Multi-bubbled geometries}
\label{ss:bubbles}

These were the first scaling, supersymmetric microstate geometries to be constructed, the techniques are now well-understood and a large range of such geometries has been constructed and analyzed  \cite{Giusto:2004kj,Bena:2005va,Berglund:2005vb,Bena:2006is,Bena:2006kb,Bena:2007kg,Bena:2007qc,Bena:2008wt,Bena:2010gg,Bena:2011dd, Bena:2013dka, Bianchi:2016bgx, Bianchi:2017bxl, Heidmann:2017cxt, Bena:2017fvm, Avila:2017pwi,Tyukov:2018ypq,Warner:2019jll}. The bubbling geometries that preserve a tri-holomorphic $U(1)$ isometry can be compactified to four dimensions, and are related to a special family of four-dimensional multi-center black holes \cite{Denef:2000nb,Denef:2002ru, Bates:2003vx,Balasubramanian:2006gi}.  As has been noted, while these geometries can have large classical moduli spaces, they are nowhere near large enough to account for the entropy of black holes. It has also been possible to construct bubbling geometries corresponding to non-supersymmetric extremal black holes, both almost-BPS ones \cite{
Goldstein:2008fq,
Bena:2009ev,
Bena:2009en,
Bobev:2009kn,
DallAgata:2010srl,
Vasilakis:2011ki}, and Kerr-like ones \cite{Heidmann:2018mtx}.

A recent breakthrough \cite{Heidmann:2021cms,Bah:2022yji}\footnote{Building on earlier work in \cite{Bah:2020ogh,Bah:2020pdz,Bah:2021owp,Bah:2021rki, Bah:2021irr}.} has led to a systematic construction of large families of {\em non-BPS non-extremal}, multi-bubbled microstate geometries.  To solve the equations of motion, rather than the much simpler BPS equations, one must simplify the problem by imposing enough symmetry to reduce the equations to a two-dimensional problem.  Despite this specialization, there are still remarkably rich families of non-BPS microstate geometries, and there are even microstate geometries with vanishing total charge.

The results in \cite{Heidmann:2021cms} also make a very important technical advance.  These solutions  have incorporated a much broader range of topological cycles and fluxes than previous microstate geometries, and this has opened the way to yet larger families of such geometries, sampling ever more extensive parts of the black-hole phase space. 

One of the early triumphs of scaling microstate geometries was to show how quantization of the moduli, and the angular momentum, would cut off the scaling at a maximal depth \cite{deBoer:2008zn, Bena:2007qc}, and that, at this depth, the simplest excitations have an energy gap of order the mass gap of the maximally-twisted sector of the CFT \cite{Bena:2006kb,deBoer:2008zn}.    Thus scaling microstate geometries access the ``typical sector'', where most of the microstructure lies.  A major open problem is how to access coherent twisted-sector excitations of microstate geometries.   One might argue that such things are intrinsically stringy, but even so, there should be coherent combinations of such excitations that can be captured by supergravity.  

It is possible that we already have the answer to this conundrum:  such coherent condensates might be dual to multi-bubbled solutions.  There is some circumstantial evidence for this based on fractional spectral flows \cite{Giusto:2012yz,Shigemori:2021pir}, but more work needs to be done before this becomes convincing. There are  thus two huge sectors on each side of the holographic duality that are, as yet, not incorporated into the dictionary.  It would be ``economical'' if these two unknowns were ``the same''.

\subsection{Superstrata and microstrata}
\label{ss:strata}

Superstrata   and microstrata are microstate geometries for the D1-D5 system, and have only one non-trivial cycle, the $\IS^3$, of the AdS$_3\times\IS^3$ of the near brane limit.  The background geometry is rendered smooth by adding angular momentum and KK dipole charge, thus creating a cap in the infrared that resembles the global AdS$_3\times\IS^3$ cap of the supertube (Lunin-Mathur) geometries.  Superstrata and microstrata are obtained by adding specific families momentum excitations to this system.  If the momentum excitations are purely left-moving, then the right moving sector remains in its ground state and one obtains \nBPS{8} supersymmetric {\it superstrata}.  Microstrata involve specific families of both left-moving and right-moving momenta, and thus all supersymmetry is broken.

The construction of superstrata in IIB supergravity is now a well-established algorithm in space-times that are asymptotic to  AdS$_3\times\IS^3 \times  \IT^4$ (or  AdS$_3\times\IS^3 \times  K3$)%
\footnote{The choice of  $\IT^4$  or $K3$ modifies the spectrum of the supergravity theory, but superstrata can be constructed  in a ``universal sector'' that is common to both compactications.   We will continue to refer to the $\IT^4$ compactification, with the understanding that we are working in this universal sector and the results extend to $K3$.}  \cite{Bena:2016ypk,Bena:2017xbt,Heidmann:2019zws}  
 or asymptotic to  $\IR^{4,1} \times  \IT^5$ \cite{Bena:2017xbt}.   The most general superstratum is governed by shape modes that are arbitrary functions of three variables.  Holographically, these correspond to coherent states of precisely-known CFT excitations  called the ``supergraviton gas'', which is generated by the lowest, non-trivial modes in the untwisted sector of the CFT.  Once again, while this system has many independent excitations, it describes a parametrically small subset of all the possible microstates \cite{Shigemori:2019orj,Mayerson:2020acj}.  Orbifolds of the geometry can access some of the twisted-sector states \cite{Bena:2016agb}, but again only a parametrically small subset.  Despite this, it is important to note that superstrata include some families of states that contribute to the index computed in \cite{Strominger:1996sh} (see, for instance,~\cite{Mayerson:2020acj}),
and so superstrata provide an important first step in answering the question: What are the Strominger-Vafa states when gravity is turned on?%
\footnote{There is an important subtlety underlying this observation: Index computations {\it are} sensitive to the compactifcation geometry, and get different results for $\IT^4$ and $K3$; the former has additional fermion zero modes which cause additional cancellations~\cite{Maldacena:1999bp}.  Supertstrata count all states and are insensitive to whether the compactifying manifold is $\IT^4$, an orbifold of $\IT^4$ or $K3$. Nevertheless, the counting of superstratum states in~\cite{Mayerson:2020acj} suggests that superstrata are indeed capturing the supergravity descriptions of some of the index states. }

Despite the relatively limited phase space of superstrata, they have become a major workhorse for probing and testing ideas surrounding microstate geometries, precision holography and fuzzballs \cite{Kanitscheider:2006zf,Kanitscheider:2007wq,Taylor:2007hs,Giusto:2015dfa,Bombini:2017sge,Giusto:2019qig, Tormo:2019yus, Rawash:2021pik}.  Some of this will be discussed in Section~\ref{Sect:Behavior}.  

A recent breakthrough \cite{Ganchev:2021pgs, Ganchev:2021ewa} resulted  in the construction of the first microstrata: supertubes carrying coherent combinations of both left- and right-moving  momentum excitations.  These geometries  were obtained using techniques of consistent truncation in space-times that are asymptotic to  AdS$_3\times\IS^3 \times  \IT^4$ \cite{Samtleben:2019zrh,Mayerson:2020tcl,Houppe:2020oqp}.  Microstrata  represent non-BPS momentum waves in equilibrium with a gas of gravitons in the ``box'' created by AdS$_3$.  When coupled to flat space, it is expected that microstrata will decay into a coherent analog of Hawking radiation.   The construction of microstrata is very new, and suggests many  follow-up projects, both in gravity and in the holographic CFT.  For the present, it is extremely gratifying that such solutions exist and that supersymmetry breaking in superstrata does not result in  collapse, as some had conjectured.

One of the other apparent limitations of superstrata is the non-zero angular momentum needed to render a smooth background. This is inherited from the fact that the supertube needs angular momentum to be smooth.\footnote{Bubbling geometries do not have this problem, as their angular momentum can be changed freely by moving the bubbles around.}   In the superstratum, one can add lots of momentum and in so doing, the angular momentum can be made arbitrarily small; however, it cannot be set to without creating a singularity.  According to the fuzzball principle, such a singular corner of the moduli space must be resolvable, and indeed it is.  The most trivial solution would be to add more bubbles and use them to cancel the angular momentum of the superstratum.  A more interesting possibility is to create superstrata with momentum carriers on the $\IT^4$ or $\IT^5$, without turning on angular momentum.  A first step in this direction has been made in \cite{Bena:2022sge,Ceplak:2022wri}.

\subsection{Sampling and supporting microstructure}
\label{ss:sampling}

A common question is: How many fuzzball states have been constructed?

At one level, this looks like a strange question. We do not construct all possible states of planets to know that planets do not pose an information paradox. We may imagine a special state of a planet as a regular crystal, and a different state in the form of an ideal gas, but we would not seek to write out all the possible quantum states of all possible planets. 

With black holes, the information paradox arose because extensive efforts failed to find `hair', and the standard solution with vacuum at the horizon led to the information puzzle. The goal of the fuzzball and microstate geometry constructions is to show how the no-hair theorems are broken in string theory, by constructing examples of specific microstates of the black hole, and noting that these have no horizon. With the large class of fuzzball solutions and microstate geometries that have been constructed, it is already clear that the no-hair theorems are broken in string theory, and this points toward a resolution of the paradox.

But one could still worry that the  solutions found so far might be some very special states of the theory, and other states could look like the traditional black hole with horizon; this would require us  to look for some alternative resolution to the information puzzle. This is where the dual CFT description becomes very useful. At least in the weak-coupling limit, we can list all the microstates of the system.  We can then identify various fuzzball solutions as strong coupling limits of these states, thereby getting an idea of what part of the space of states we are sampling in our constructions. As we will discuss below, the constructions now sample so many corners of the space of microstates that no longer seems reasonable to argue that other microstates will have traditional horizons.

An example of a system where one can show that all CFT ground states correspond to horizonless configurations, and that the spherically-symmetric (small) black hole solution with a horizon is not dual to any pure state of the CFT, arises in the D1-D5 system. The two-charge BPS ground states of the CFT are dual to supertubes of arbitrary shapes, and the quantization the moduli space of these supertubes reproduces the entropy of the two-charge black hole\cite{Palmer:2004gu, Rychkov:2005ji}. When the supertubes are large, their fully back-reacted supergravity solutions are smooth and have low curvature, and are thus microstate geometries. However, the typical states of the D1-D5 system are dual to supertubes whose curvature is large and cannot be described using supergravity \cite{Lunin:2002iz,Chen:2014loa}, but can be approached as limits of smooth supergravity solutions, and can also be constructed via world-sheet methods \cite{Martinec:2020gkv}.

The two-charge black hole has string-scale curvature at the horizon, but taking into account an infinite tower of string-theory corrections \cite{Dabholkar:2004yr} one can show that it develops a small horizon. The Wald entropy of this small black hole matches perfectly the entropy of the dual CFT, and hence we can argue that for the two-charge system the fuzzball proposal is realized: the microstates counted by the black-hole entropy are solutions without a horizon.\footnote{Strictly speaking the two-charge small black hole exists in a duality frame where the two-charge microstates are not smooth supergravity microstate geometries but rather microstate solutions \cite{Sen:2009bm}. However, this does not contradict the fuzzball proposal \cite{Mathur:2018tib}}. The fuzzball idea proposes that, because of the fractionation effects discussed above, even when the horizon is much larger than string-scale, it is supplanted by horizonless microstructure.
 
We do not know how to construct all three-charge extremal solutions in closed form, but it is natural to ask if these could also all be approached through solutions in classical supergravity. We do not know the answer to this question, but condensates of strings are known to generate microstate geometries, which are again described by classical supergravity. 

Another possibility is that the  microstate geometries provide ``coherent islands'' in a vast family of stringy microstates.  The question then becomes how densely spread are these coherent islands.  Could they sample enough of the phase space to capture the thermodynamics of all the microstructure, in much same way that simple, classical kinetic models described the thermodynamics of ideal gases long before quantum mechanics came into existence?  Given the incredible range and diversity of known superstrata, and the fact that finding more of them seems to be a technical, rather than a fundamental limitation, gives many of us cause to think that is is a real possibility.

Several classes of non-BPS microstate geometries have been constructed, while other non-BPS states have been constructed using stringy excitations.  Strings in the pp-wave limit give an instance of a non-BPS gravity solution where the dual CFT state can be identified in a controlled approximation \cite{Gava:2002xb}.  More generally, we can place both BPS and non-BPS metastable supertubes in microstate geometries. It is possible that coherent states of such stringy excitations give classical supergravity solutions (see, for example, \cite{Bena:2008dw,Bena:2011fc,Bena:2012zi}), but it is also possible that the generic state cannot be approached through classical solutions, and generalizations of the foregoing quantum excitations would have to be understood in a systematic way.

Overall, we have been able to reach a huge number of points in  the phase space of black holes, by taking microstate geometries and  studying the behavior of gases of strings and supertubes  in such backgrounds. {\it Every} string theory state that has been constructed has turned out to be a fuzzball: a state which has no horizon.  Given these results
it does not appear reasonable that the black hole has any states that are {\it not} fuzzballs.

\section{String world-sheets and fuzzballs}
\label{Sect:WS}

The passage from microstate geometries to generic fuzzballs requires string theory rather than its massless limit, supergravity.  The first step is to study perturbative string theory effects in non-trivial, but simple microstate geometries.  
The world-sheet formalism yields a perturbative expansion about such a background, but can handle situations in which the supergravity approximation breaks down.  In particular, we gain insights into the spectrum of stringy excitations around the background; we can compute processes involving string-scale momentum transfers among the excitations; and we can investigate  the properties of backgrounds with string-scale curvature.  Indeed, stringy excitations can play an essential role in resolving some of the singular corners of the supergravity moduli space.  For example, the world-sheet theory resolves the locations of individual five-branes, and exhibits the singularities that arise when five-branes collide.  These singularities are resolved by additional stringy effects, such as the appearance of new light degrees of freedom (``W-branes'') at five-brane intersections, revealing the sorts of fractionation effects that are key to understanding black hole entropy and generic fuzzball states.

\subsection{The world-sheet sigma model }

The string world-sheet action is a two-dimensional non-linear sigma model with a Wess-Zumino term.  The precise action is determined by ``pulling back'' certain fields of the background onto  the world-sheet.  Remarkably enough, the string dynamics is  exactly solvable in an  $AdS_3\times \IS^3\times\IT^4$ geometry (because it is a tensor product of group manifolds)\footnote{The WZ terms on $AdS_3$ and $\IS^3$ describe electric and magnetic three-form field strengths sourced by the fundamental strings and NS5-branes whose near-horizon geometry the 
$AdS_3\times \IS^3\times\IT^4$ solution is.}.  At the quantum level, the underlying current-algebra symmetries are sufficiently constraining that their Ward identities determine all the correlation functions~\cite{%
Teschner:1997ft,
Giveon:2001up,
Maldacena:2001km},  
and so perturbative string dynamics about vacuum AdS is in principle exactly solvable~\cite{%
Giveon:1998ns,
Kutasov:1999xu,
Maldacena:2000hw,
Maldacena:2001km}.

Moreover, the world-sheet non-linear sigma model on the ground-state backgrounds of~\cite{Lunin:2001fv} has sufficient supersymmetry ($\cN\!=\!(4,4)$ in two dimensions) to guarantee that the action does not receive quantum corrections~\cite{Horowitz:1994ei,Horowitz:1994rf}.  As a consequence, all these backgrounds are exact solutions to string theory, even if we can't solve them exactly. 
Fortunately, many of the  families of solutions most studied in the literature~-- the ``round supertubes'' of~\cite{%
Lunin:2001fv,
Giusto:2004id,
Jejjala:2005yu,
Giusto:2012yz,
Chakrabarty:2015foa} 
whose geometry is the quotient $(AdS_3\times \IS^3)/\IZ_k\times\IT^4$~-- admit an exact world-sheet description in terms of a group coset sigma model (a gauged Wess-Zumino-Witten model)~\cite{Martinec:2017ztd}.  The curvature singularity at the $\IZ_k$ orbifold point is resolved by the world-sheet dynamics, and in particular, concerns raised about the stability of these geometries in~\cite{Eperon:2016cdd,Marolf:2016nwu} are resolved within perturbative string theory~\cite{Martinec:2020gkv}.

The spectrum of string excitations and the light (but possibly stringy) operator correlation functions about the round supertube can thus be worked out exactly in terms of the underlying current algebra structure~\cite{Martinec:2018nco,Martinec:2020gkv}.%
\footnote{The consistency of the string spectrum is intimately linked to smoothness and absence of CTCs in the corresponding supergravity backgrounds~\cite{Bufalini:2021ndn}.}
In particular, one finds the BPS vertex operators that effect transitions among the supersymmetric ground states, and their holographic map to the corresponding CFT operators.  Exponentiating these vertex operators into the action makes a coherent deformation of the background which preserves exact superconformal invariance, and exhibits in the bulk string theory the same structure seen in the dual CFT description of these ground states: that they are coherent states of supergraviton excitations.

\subsection{Stringy resolutions of two-charge fuzzball structure}

The family of ground-state sigma models exhibits the cap at finite redshift seen in~\cite{Lunin:2001fv}, but one can now analyze the typically stringy near-source structure.  In regions of strong curvature, there can be strong mixing between string-theory modes, such that the metric geometry of the non-linear sigma model can have a dual presentation in terms of a superpotential~\cite{%
Martinec:1988zu,
Vafa:1988uu,
Ooguri:1995wj,
FZZref,
Giveon:1999px}.  
This superpotential represents the winding-string condensate bound to the five-brane background, as was first observed in a closely-related context (see~\cite{Giveon:2015cma,Giveon:2016dxe,Martinec:2020gkv} for a discussion).
The key point is that the shape of the supertube directly codes the form of the superpotential, and the stringy effects that occur at strong curvature.
A satisfying picture emerges in which singularities in the geometry are mirrored by degenerations of the superpotential~\cite{Martinec:2020gkv} for a wide class of supertubes.  The world-sheet theory develops a strong-coupling singularity when this occurs, in which strands of the five-brane source self-intersect; these self-intersections result in the breakdown of the world-sheet perturbation theory.  This breakdown does not mean that string theory is failing, rather just that certain singularities in gauge theory and string theory are only resolved by strong-coupling dynamics (see for example~\cite{Seiberg:1994rs,Strominger:1995cz,Katz:1996ht,Seiberg:1996vs}).

What are the strong-coupling phenomena here?  The non-abelian ``tensionless'' (more precisely, low-tension) string theory that governs the internal dynamics on a stack of coincident five-branes~\cite{Dijkgraaf:1997ku,Seiberg:1997zk,Giveon:1999px,Giveon:1999tq}, known as ``little string theory'' (see~\cite{Aharony:1999ks,Kutasov:2001uf} for reviews).  The five-branes in a typical supertube source are separated in their transverse space, and as a result the bubbled cycles in the geometry have finite size, and the geometry is smooth (or, at worst, has stringy but finite curvature).  But there are regions of the source configuration space where five-branes collide, and the cycles shrink away. 

Pairs of (or multiple) coincident five-branes lead to a partial de-confinement of non-abelian five-brane excitations, the little strings.  These are none other than the fractionated strings responsible for black hole entropy~-- indeed, the entropy~\eqref{Sbtz} can be regarded as a limit of the Hagedorn entropy of little strings~\cite{Martinec:2019wzw}.  
One sees in the bulk dynamics deep in the AdS throat the emergence of features of the gauge theory underlying the black-hole phase and its thermodynamics~\cite{Martinec:2019wzw,Martinec:2020gkv}.   As has been seen in other contexts (\eg~\cite{Strominger:1995cz,Witten:1995zh}), and as another expression of the  {\it ``fuzzball principle"},  whenever a geometrical singularity seems to develop, new light degrees of freedom emerge to resolve it.  Here, the apparent singularity is a two-charge black-hole extremal horizon, and the new light degrees of freedom are the non-abelian ``little strings'' that arise when five-branes coincide~\cite{Dijkgraaf:1997ku,Seiberg:1997zk,Giveon:1999px,Giveon:1999tq}.

\subsection{Stringy explorations of more general microstate geometries}

Deformations of the world-sheet sigma model preserving the maximal $\cN=(4,4)$ supersymmetry are thus rather well understood.   More generally, one may consider deformations that preserve less supersymmetry.  For instance, superstratum backgrounds preserve only $\cN=(0,4)$ supersymmetry, and in the worldsheet formalism result from condensing supergraviton vertex operators preserving only this reduced supersymmetry.  A subset of these may be amenable to world-sheet analysis as $\cN=(0,4)$ non-linear sigma models, 
for which non-renormalization theorems are almost as powerful in controlling the quantum corrections.  Of particular interest are possible generalizations of the duality between geometry and superpotentials discussed above, 
which could describe the near-source structure of a class of superstrata and their degenerations as features of a corresponding superpotential.  The fact that the dual superpotential in the two-charge background codes the stringy near-source structure and the means by which string perturbation theory breaks down leads to the hope that a similar analysis in the $\cN=(0,4)$ theory might help characterize the transitions between superstrata microstate geometries and generic fuzzballs.  The superstrata discussed above are built on a foundation of supersymmetric ground states by coherently adding chiral momentum excitations.  One might expect the degenerations of the ground states in which five-brane sources self-intersect, extend to superstrata in some manner.  Such behavior would again signal a transition in which a rather coherent microstate geometry, which can lie deep within the ensemble of black hole microstates, scrambles into more generic fuzzball states.

The world-sheet analysis of these degenerating two-charge
supergravity solutions suggests that they are rather non-generic in the state space, but they are useful launching points for an analysis of the sorts of strong coupling effects that occur in black holes.  For instance, if we supply some energy above a smooth supersymmetric ground state, it will start wandering the large phase space of source configurations until the source five-branes in the background self-intersect, at which point little string excitations will appear and bind the five-brane strands together at the intersection.  The background five-branes and the little string excitations supported on them are a description of the more generic fuzzball that is emerging near extremality.  We are now seeing this structure emerge in the bulk description in addition to the weakly-coupled regime of the dual CFT.

\section{Replication of, and deviations from, black-hole behavior }
\label{Sect:Behavior}

Much of the physics of black holes appears to come from the existence of an event horizon, and so, if one is to replace objects with horizons by microstate geometries, one must address the question: ``How do these horizonless solutions reproduce the essential physics of a black hole?"   For example, a probe particle falling into a black hole appears to vanish behind an event horizon, while a particle falling into a microstate geometry appears to bounce back in a finite time.  In this Section we illustrate how some of the known black-hole properties  emerge from horizonless microstate geometries. Furthermore, we will argue that, once seen in the context of the AdS-CFT correspondence, some of the differences between microstate geometries and classical black-hole geometries are features and not bugs.

\subsection{Red shifts and energy gaps}

Solving the wave equation of an arbitrary field in a classical black-hole geometry only gives rise to solutions that decay with time, and this is consistent with black hole uniqueness and with the expectation that black holes are completely absorbing objects. Hence, black holes support no standing waves. Furthermore, they have an infinite redshift between the horizon and infinity. In contrast, microstate geometries support an infinite tower of linearized perturbations, with a finite energy gap. Moreover, in microstate geometries the maximal redshift between asymptotic infinity and the horizon-replacing cap is typically very large, but finite.

For special families of microstate geometries there exists a ``scaling limit'' that approaches an extremal black-hole solution arbitrarily closely, and in this limit the redshift does become infinite and the energy gap goes to zero, collapsing the whole tower of excitations.  The  limit of these scaling solutions has an infinitely-long black-hole-like throat, and the black-hole microstate geometries can approach continuously the black-hole solution \cite{Bena:2006kb}.  Nevertheless, it has been argued in \cite{Bena:2007qc,deBoer:2008zn,Li:2021gbg,Li:2021utg} that this intuition, which comes from studying classical solutions, is not correct, and quantum effects can prevent the collapse of a bubbling microstate geometries into a black hole.

One of the early successes of the microstate geometry program is the exact match between the mass gap computed in certain asymptotically-AdS supergravity solutions, and the mass gap one expects of typical supersymmetric ground states of the D1-D5 CFT \cite{Bena:2006kb,Bena:2007qc,deBoer:2008zn,deBoer:2009un}. It was known for a while that both these gaps are of order $({N_1 N_5 R})^{-1}$ \cite{Bena:2006kb, Tyukov:2017uig}, but solving analytically the separable wave equation in a superstratum solution \cite{Bena:2017upb} revealed that even the numerical coefficient matches precisely that of the CFT  \cite{Bena:2018bbd}.

\subsection{Thermalization, Tidal Scrambling and Information Recovery}

One can also compute Green functions in the microstate geometries dual to certain pure states of the D1-D5 CFT, and compare them to the ``thermal-average'' Green function computed using the classical black-hole solution~\cite{Bena:2019azk}. This ``thermal-average'' Green function exhibits an exponential decay, consistent with the fact the classical black-hole solution is a purely absorptive object. In contrast, in a capped horizon-less geometry the Green function computed in the {\em linearized supergravity approximation} mimics the black-hole absorptive behavior for a long time but, at a time of order the inverse of the mass gap, has an ``echo''.

In chaotic systems, the Green function exhibits a dip-ramp-plateau behavior as a function of the time separation.  In holographic systems, this behavior is seen both at finite \cite{Maldacena:2015waa} and at zero temperature \cite{Craps:2020ahu,Craps:2021bmz}.  
In microstate geometries, deviation from thermal decay is expected to appear at times of order the inverse mass gap, which differs between extremal and non-extremal systems.  As argued in \cite{Bena:2019azk}, the echo seen in the microstate geometry Green function calculation is consistent with this expectation. 

Linearized analysis in microstate geometries gives  interesting physics, whose difference from black holes might be of experimental relevance. In addition to reflections and echoes, a stationary wave at the bottom of an asymptotically-flat microstate geometry will decay  \cite{Chowdhury:2007jx,Chakrabarty:2015foa,Chakrabarty:2019ujg,Bena:2020yii}.  One can argue that the quasi-normal modes characterizing this decay correspond to the ring-down of the classical black hole solutions. Furthermore, by doing the Green-function calculation in asymptotically-flat space and carefully disentangling the decaying modes, one can compute tidal Love numbers for the microstate geometries \cite{Love-to-appear}.

These calculations are important \cite{Mayerson:2020tpn,Mayerson:2022yoc}, since ring-down frequencies, echoes and tidal Love numbers can be extracted from gravitational-wave data. These calculations can be done both in superstrata \cite{Bena:2020yii,Love-to-appear}, where the wave equation for scalar fields is separable \cite{Bena:2017upb}, and in more complicated BPS bubbling geometries where no such separability exists \cite{Bianchi:2021mft, Ikeda:2021uvc}. While these calculations are done in microstate geometries of black holes that are far from real-world black holes, some of the physics they reveal could be universal, and characterize horizon-scale structure of generic black holes. 

Some of these calculations will be greatly modified by another remarkable feature of microstate-geometry physics: {\em tidal scrambling}
\cite{Tyukov:2017uig, Bena:2018mpb, Bena:2020iyw,Martinec:2020cml,Ceplak:2021kgl}.

Scaling microstate geometries  have long throats, and  infalling probes accelerate to ultra-relativistic speeds. These probes experience strong tidal forces, even at large distances  from the structure replacing the horizon \cite{Tyukov:2017uig, Bena:2018mpb, Bena:2020iyw} because the blue-shift hugely amplifies the tiny differences between the black-hole and microstate geometry. As a result of these tidal forces, the supergravity approximation can break down.   In string theory, particle probes are actually strings whose oscillator modes are in their ground state; when the tidal forces exceed the string scale, these oscillator modes become excited \cite{Martinec:2020cml,Ceplak:2021kgl}. Such excitations give the string a non-zero mass, and the energy comes from the kinetic energy of the probe.   As a result, the probe can no longer return to infinity to produce the echo of \cite{Bena:2019azk}.  Instead, these massive strings are trapped; they decay, and while some of the massless fields produced by the decay could escape to infinity, others will remain at the bottom of the superstratum and initiate a thermalization process.  Thus, in solutions with a long throat, {\it tidal scrambling leads to the  trapping of matter, and the onset of thermalization.}   This means that  the outcome of supergravity calculations, like that of \cite{Bena:2019azk},  are modified in string theory.  Indeed, the stringy dynamics  means that microstate geometries approach the purely absorptive behavior of black holes, and the shape of the signal will not be a perfect echo, but rather a series of small uncorrelated signals, similar to what was computed in the D1-D5 CFT in \cite{Craps:2020ahu,Craps:2021bmz}.

Most of these calculations are based on probes, and so the background is fixed.  Such scattering calculations  ignore the back-reaction and the possibility of the microstate geometry evolving into other background geometries nearby in phase space. This assumption is justified if scattering off the background is predominantly elastic.
However, we know that both superstrata and multi-bubbled  solutions admit (classically) a very large number of massless deformations. This property is shared by two-charge D1-D5 microstates, which  are classically parameterized by continuous source profile functions \cite{Lunin:2001fv, Lunin:2002iz, Mateos:2001qs, Emparan:2001ux}. It is only after quantization of the phase space of these supertubes \cite{Palmer:2004gu, Rychkov:2005ji}  that one realizes that the dimension of their moduli space of deformations is not infinite but rather of order $N_1 N_5$, and one obtains the entropy of the two-charge system. 
A similar argument can be made that the dimension of the moduli space of deformations of superstrata is of order $N_1 N_5$ \cite{Bena:2011uw, Bena:2014qxa, Shigemori:2019orj, Mayerson:2020acj}, and hence an incoming wave will not leave the superstratum microstate geometry unscathed, but rather cause motion on the phase space into nearby microstate geometries.\footnote{A similar mechanism was pointed out for two-charge systems in \cite{Marolf:2016nwu}.}  Similarly, bubbling microstate geometries have a moduli space that is classically continuous \cite{Bena:2005va,Berglund:2005vb,Bates:2003vx}; its quantization \cite{deBoer:2008zn} reveals however that it can only support a finite number of states. Furthermore, since the mass gap corresponding to excitations above a given long-throat superstratum or bubbling geometry is of order $1/N_1 N_5 R$, an incoming particle can also excite the superstratum into a non-supersymmetric configuration with more mass than charge (corresponding to a microstate of a non-extremal black hole) and be completely absorbed.

To summarize, we expect microstate geometries and fuzzballs to absorb infalling particles, by a combination of {\it at least} three phenomena: tidal scrambling, phase-space motion and exciting of low-mass-gap non-BPS excitations. The precise interplay between these phenomena, and their relative importance as one moves from BPS to non-BPS fuzzballs and microstate geometries is an important subject for future investigation.

\subsection{Multipole moments, shadows and echoes}

One important feature that distinguishes microstate geometries and fuzzballs from the black hole with the same charges are the multipole modes.  All known microstate geometries\footnote{This is probably just a technical limitation: recent work \cite{Bena:2022sge,Ceplak:2022wri} suggests that there may well exist spherically symmetric microstate geometries}  break the spherical invariance of the black-hole solution, and this breaking can be parameterized by the presence of non-trivial multipole moments \cite{Bena:2020see,Bianchi:2020bxa,Bena:2020uup, Bianchi:2020miz,  Bah:2021jno, Fransen:2022jtw, Loutrel:2022ant}. Understanding the magnitude of these deviations  from the black hole is an important question, especially given the expectation that LISA measurements of extreme mass-ratio inspirals (EMRI) will place strong bounds on a considerable set of such moments \cite{Babak:2017tow,Fransen:2022jtw, Loutrel:2022ant}.

Extremal microstate geometries also have a long throat in which, as the cap descends to deeper and deeper red-shifts, they increasingly resemble the corresponding black hole. 
The inverse of this characteristic redshift serves as a scaling parameter; taking it to zero, the multipole moments of the BPS microstate geometries resemble more and more closely those of the BPS black hole \cite{Bena:2020see,Bianchi:2020bxa,Bena:2020uup, Bianchi:2020miz}. Hence, it appears naively that all information about horizon-scale structure is lost. However, one can define dimensionless ratios of vanishing multipole moments \cite{Bena:2020see,Bena:2020uup}, which remain finite as the microstate geometries approach the black hole, and which parametrize deformations of the black hole. 

There are almost-BPS black holes \cite{Bah:2021jno} that, unlike  their BPS counterparts, can have finite angular momentum \cite{Bena:2009ev}, and so look closer to Kerr black holes.  These black hole have infinite towers of non-trivial multipole moments  \cite{Bah:2021jno}, and the microstate geometries of these black holes \cite{Goldstein:2008fq,Bena:2009ev,Bena:2009en} have multipole moments that generically differ from those of the black hole \cite{Bah:2021jno}. Once again, these differences are proportional to the scaling parameter \cite{Bah:2021jno}. It is unclear whether the coincidence of the fuzzball multipole moments to those of the black hole in the scaling limit is an artifact of extremal black hole, or is rather a universal feature that will apply also to microstates geometries for the Schwarzschild and Kerr black holes. The obvious next step  is to build upon the results of \cite{Heidmann:2021cms,Bah:2022yji} and construct and such  microstate geometries and analyze their multipole moments.

Shadows provide another experimentally-measurable quantity that could distinguish black holes from microstate geometries.  Horizonless solutions trap light for long periods near the would-be horizon, and this produces a shadow that differs from that of a black hole, where all light is absorbed \cite{Bianchi:2018kzy,Bianchi:2020des,Bianchi:2020yzr,Bacchini:2021fig}.  So far, calculations have only been done for microstate geometries corresponding to extremal BPS black holes.  The results exhibit the same behavior as extremal BPS black-hole multipole moments: as one approaches the scaling point and the microstate geometry approaches closer and closer the black hole, the microstate shadow approaches the black-hole result.

One might also  see signals of microstate geometries in the ``noise'' from measurements.  It appears that variations  of the multipole moments about black-hole values may well average to zero\cite{Bah:2021jno}, contrary to some early hopes \cite{Bianchi:2020bxa}.  More interestingly, some of the ``dissipative'' quantities computed using black-hole microstate geometries, like chaos \cite{Bianchi:2020des} or tidal Love numbers \cite{Love-to-appear} are expected to differ from the corresponding black-hole quantities in a manner that does not average to zero\footnote{The tidal Love numbers of four-dimensional black holes are zero, hence this is not surprising.}.  Indeed, as one would expect from  dissipation: adding more channels makes the effect larger.  It would be interesting to understand whether these features are an artifact of using extremal black holes, or whether this is a universal feature that applies also to microstates of non-extremal black holes.

\section{Open questions and future progress}
\label{Sect:future}

In lieu of a more formal conclusion section, we will finish by outlining some of the important open problems in microstate geometries and fuzzballs

\subsection{Microstate geometries,  black-hole microstructure and superstrata}
\label{ss:futurestructure}

The construction of microstate geometries is far from complete: every year, new classes of such geometries are constructed, or conjectured to exist.  Indeed, the last year  has seen a huge range of new possibilities \cite{Ganchev:2021pgs,Ganchev:2021iwy,Heidmann:2021cms,Bah:2022yji, Bena:2022sge,Ceplak:2022wri}.

One of the most important standout problems is the holographic interpretation of multi-bubbled geometries.  The  recent results for non-BPS microstate geometries \cite{Heidmann:2021cms,Bah:2022yji} give this issue even greater urgency.   There is also the intriguing possibility of putting momentum waves on multiple bubbles and making multi-superstrata.  This is technically challenging, but not impossible if one uses the insights of \cite{Tyukov:2018ypq,Walker:2019ntz} combined with the Green functions of \cite{Page:1979ga}.  

As noted in Section \ref{ss:status-overview}, another open question is the systematic treatment of the geometric transitions that lead to bubbled solutions.  These transitions have typically been studied in low space-time dimensions with a ``passive'' compactification manifold.  However, the results of \cite{Heidmann:2021cms,Bah:2022yji} make the important step of creating transitions that are more democratic within the compactification.  There are also likely to be many new families of microstate geometry that emerge from hybridizing the original supersymmetric solutions with those of  \cite{Heidmann:2021cms,Bah:2022yji}.

On the other side of the duality, we know that most of the black-hole microstructure comes from the highly fractionated, twisted-sector states, and yet the with the exception of \cite{Bena:2016agb}, the microstate  geometries that have known holographic duals can only access the states in which the momentum is carried by non-fractionated modes. Hence, even if we believe there should exist solutions dual to states with generic coherent combinations of fractionated momentum carriers, we do not have the technical ability to construct them yet. It has been conjectured that they might be related to multi-bubbled solutions \cite{Giusto:2012yz,Shigemori:2021pir}, but this remains unknown territory.

\subsection{Non-BPS solutions}
\label{ss:nonBPS}

There have been two major breakthroughs in the systematics of constructing non-BPS microstate geometries. These results are in a very early stage and there are many directions that can be developed.  

For the non-BPS bubbled geometries  \cite{Heidmann:2021cms,Bah:2022yji}, there are many obvious questions.   What are the decay channels and how do they emit Hawking radiation? What are the families of geometries that can be constructed using the present techniques? Can one include magnetic fluxes and angular momentum?  Is there a smooth deformation that links them to their supersymmetric counterparts? Can they be thought as corresponding to the geometric transition of a system of branes ? How might one find solutions that are not axisymmetric and depend on more than two variables? 

Microstrata have been constructed in the simplest possible way using a combination of superstratum technology and the ``Q-ball trick'' \cite{Ganchev:2021pgs,Ganchev:2021ewa}. These solutions involve only a single excited non-BPS mode, and the first paper was intended as a ``proof of concept''. There are potentially huge generalizations of this class of solution using both numerics and perturbation theory.  Perhaps the most exciting extension of this work would be  the computation of the analog of Hawking radiation from such geometries by coupling them to flat space.

Microstrata and superstrata are most easily constructed in asymptotically-AdS space-times, but  for superstrata the technology for asymptotically-flat solutions is well-established  \cite{Bena:2017xbt}.  One should be able to adapt these methods to construct  asymptotically-flat microstrata.  Such solutions are expected to decay through tunneling transitions and these can be computed using the techniques developed in \cite{Bena:2020yii}. This result can also be tested against the corresponding closed-string emissions in the CFT.

Superstrata have zero temperature and a rigid spectrum, but microstrata exhibit a spectrum whose frequencies depend on  the amplitudes and dynamics of the excitations.   This ``frequency shifting'' is an essential indicator of how the spectrum of microstrata will become chaotic.  It would be extremely interesting to study this phenomenon in more complex microstrata with multiple excitations of different frequencies.

\subsection{Scrambling and transitions to fuzzballs}
\label{ss:scrambling}

Geodesic  \cite{Tyukov:2017uig, Bena:2018mpb, Bena:2020iyw,Bena:2020iyw} and string-probe \cite{Martinec:2020cml,Ceplak:2021kgl}  analyses have revealed that tidal forces become large in microstate geometries and that this results in string excitations that causes tidal trapping of probes.   This is an important first step in understanding the scrambling process in microstate geometries.  Beyond this, one would like to understand how such an infalling string scrambles into the fuzzball states in the neighborhood of such a geometry, presumably thermalizing and ultimately back-reacting into the geometry.  

Since the incoming state has very high energy, and the states in the cap have the energy of the twisted sector states, this thermalization must involve relatively few quanta scrambling into vast numbers of twisted-sector states. Computing aspects of this process would be extremely interesting. This scrambling process could also provide insights to the effective hydrodynamics of matter falling into a fuzzball.

There are a variety of excitations one could study.  At the lowest energies, there are collective motions on the moduli space of \nBPS{4} supersymmetric ground states, which correspond to supertubes.  One would like to understand how an infalling probe excites these modes. Computing these excitations in the bulk corresponds in the dual CFT to computing correlators between two heavy and one or two light fields, where the two heavy operators are different. 

As discussed in Section~\ref{Sect:WS}, there are subspaces of this configuration space where the supertube source self-intersects.  The separation of the source strands locally controls the size of a cycle in the geometry; when the source self-intersects, the cycle collapses.  Light wrapped brane excitations arise at the intersection point.  These wrapped branes are the ``W-brane'' quanta that become massless and regularize singularities in the moduli space of non-abelian vector or tensor gauge theories~\cite{Seiberg:1996vs,Strominger:1995cz,Witten:1995zh,Seiberg:1996vs}.  It is plausible that these points in configuration space are portals to the more generic fuzzballs that dominate the black hole ensemble, and so their properties and role in thermalizing microstate geometries are important avenues of investigation.

While the analysis of such degenerations is best understood for two-charge \nBPS{4} ground states, there are indications that a similar structure arises for three-charge \nBPS{8} multi-bubbled microstates~\cite{Martinec:2015pfa}.  Superstrata are built on supertubes by adding momentum-carrying excitations, and so one might expect them to also exhibit similar phenomena; however superstrata built on more general supertubes have not yet been constructed.  As mentioned in Section~\ref{Sect:WS}, one possible approach is via an analysis of $\cN=(0,4)$ world-sheet theories.

One would also like to understand the processes by which tidally trapped probes of microstate geometries \cite{Martinec:2020cml,Ceplak:2021kgl} thermalize.  There are two scales here: the curvature of the geometry controls the scattering and thermalization of vibrational modes on a given string, while the string coupling governs the rate at which excited strings split and join, as well as radiate supergravitons into the ambient geometry.

There are also  intriguing questions about the CFT dual of tidal forces and the gravitational scrambling process.  Presumably this involves a simple incoming state  decohering into a a complex quantum state composed of a vast number of twisted sector states.  The calculations in the D1-D5 CFT are very challenging but there have been some significant first steps in this direction \cite{Hampton:2019csz,Guo:2021ybz,Guo:2021gqd}.

\subsection{New horizon-scale physics}
\label{ss:newphys}

To some extent, the answer to the question of how the transition to generic fuzzballs proceeds is predicated on the answer to the questions of how the theory accesses the dominant part of the density of states, what structures underly the generic fuzzball, and what is the horizon-scale physics involved that supports the fuzzball state against collapse (beyond the qualitative Fermi Golden Rule arguments discussed above).

In the context of AdS$_3$, we have known since~\cite{Strominger:1996sh} that the entropy is accounted for by 
fractionated momentum modes on coincident D1 and D5 branes
(the D1-charge fractionates into $n_5$ pieces, so that the longest wavelength momentum mode has quantum $n_1n_5$ rather than just $n_1$).
We have now seen glimmers of that structure starting to appear in the {\it bulk} dynamics in particular limits~\cite{Martinec:2019wzw,Martinec:2020gkv}, but the analysis breaks down when strong-coupling dynamics arises.  Nevertheless, often perturbative string theory exhibits a harbinger of the sort of physics involved, for instance in the quantum spreading of a string when it is shocked or otherwise probed on short time scales~\cite{Susskind:1993aa,Dodelson:2015toa,Dodelson:2015uoa,Dodelson:2017hyu}, or the fuzzball behavior of self-gravitating strings below the correspondence transition~\cite{Horowitz:1996nw,Horowitz:1997jc,Chen:2021dsw}.

\subsection{Broader applications of fuzzballs and microstate geometries}
\label{ss:Interconnections}

We have outlined some of the important threads for future research within microstate geometries and fuzzballs.   There are also very interesting questions about how these programs connect to other areas of research on black-hole microstructure.  Indeed,  as we have argued, horizon-scale microstructure can only be gravitationally supported by microstate geometries, and so this means that microstate geometries will always have a core role in any semi-classical discussion of black-hole microstructure.

Some of the other microstructure programs focus on dynamics, index theory and Euclidean saddle points of hugely simplified models, like JT gravity.  Some of these investigations seem to involve ``toy models'' and simple analogs of black holes, while others seem to be accessing and counting the microstates of supersymmetric black holes that can be resolved into microstate geometries. (See, for example, \cite{Jafferis:2021ywg,Iliesiu:2021are}.)  There are also information-theory arguments \cite{Hayden:2020vyo}, state-counting arguments \cite{Denef:2007yt} and generic CFT arguments \cite{Kraus:2016nwo} that the states whose counting gives the black-hole entropy do not have a horizon when gravity is turned on, and are hence, {\em by definition}, fuzzballs. It would be very interesting to explore the connections between all these very different approaches to microstructure.

Microstate geometries can also help elucidate the AdS$_2$-CFT$_1$ correspondence. One can construct huge families of supersymmetric horizonless solutions that have Poincar\'e AdS$_2$ asymptotics \cite{Bena:2018bbd}, and are therefore dual to supersymmetric ground states of the CFT$_1$. These ground states break conformal invariance in the infrared and, by the usual holographic dictionary, have a plethora of operators which acquire non-trivial vacuum expectation values. This indicates that the CFT$_1$'s dual to the AdS$_2$ near-horizon regions of the extremal black holes of string theory are not  topological theories, but have non-trivial dynamics. This can also be used to show that these CFT$_1$'s do not have conformally-invariant ground states\footnote{We thank Miguel Paulos and Dalimil Mazac for discussions on this topic.}.

Going beyond the original focus on black holes, the results of the microstate geometry program have had a broader impact in the development of AdS/CFT.  For example, there was an open problem in calculating four-point functions  using Witten diagrams for AdS$_3$/CFT$_2$ \cite{DHoker:1999mqo}.  It took 20 years to solve this problem and this was first achieved, not through computing Witten diagrams, but by taking the light limit of a HHLL correlator computed in a microstate geometry \cite{Giusto:2018ovt}.  This, in turn, led to the development of new techniques for computing Witten diagrams in three dimensions and these were then used to compute four point functions  \cite{Rastelli:2019gtj}.  This work also revealed a hidden six-dimensional conformal invariance that controlled correlators of tensor multiplets.  However,  the supergraviton correlators appeared to be naively incompatible with this conformal symmetry. Once again, further insights came from microstate geometries  \cite{Giusto:2019pxc,Giusto:2020neo}, where it was shown how the hidden superconformal symmetry extended to the supergraviton multiplet.

It  therefore seems that there are some very significant  possibilities for future research in forging connections between microstate geometries and other areas of high energy physics, and especially with other descriptions of black-hole microstructure.


\subsection{Outlook }
\label{ss:Outlook}

The fuzzball  and  microstate geometry programs have been growing and developing for over two decades and
have made substantial progress in understanding the horizon structure of black holes and resolving the black-hole information paradox.  

The foundation of this work lies in a rich circle of ideas within string theory, and is supported by detailed, and sometimes challenging, computations.  Each new result  suggests further computations to test, or build out, the overall picture of black-hole microstructure, adding to the forward momentum of the research.  One of the core strengths of the  fuzzball and microstate geometry programs is that they lie at the intersection of many different areas of fundamental physics: field theory, string theory, supergravity, GR and quantum information theory.  Each of these perspectives informs the other and many advances have been propelled by such symbioses.

As we have tried to describe in this review, the fuzzball  and  microstate geometry programs provide one of the best approaches to understanding the physics black-hole microstructure.  Based on the current level of activity, and the progress to date, these programs will remain a vibrant and exciting area of future research.  There remain many open questions and avenues to explore, especially in taking what we have learned so far and using it to inform our knowledge of astrophysical black holes.

\section*{Acknowledgments}
\vspace{-2mm}
We would like to thank our many collaborators and colleagues whose work and insights enabled us to see a little further.  We would like to thank Rodolfo Russo for explaining the back story of the AdS$_3$/CFT$_2$ correlators.   
The work of NW is supported in part by the DOE grant DE-SC0011687. 
The work of EJM is supported in part by DOE grant DE-SC0009924.
The work of  IB and NW is supported in part by the ERC Grants 787320 - QBH Structure and 772408 - Stringlandscape.
The work of  SDM  is supported in part  by DOE grant DE-SC0011726.

\appendix

\section{Some common misconceptions about the information problem and its potential resolution}
\label{App:A}

The small-corrections theorem  \cite{Mathur:2009hf}   tells us that to resolve the information problem one must give up (at least) one of three things
\begin{enumerate}[\it (i)]
\item Smooth horizons for black holes
\item Unitarity of quantum mechanics in the presence of black holes
\item Locality of quantum mechanics  in the presence of black holes
\end{enumerate}
The fuzzball paradigm  embraces a violation of   option (i) and then addresses  the gravitational and quantum description of horizon-scale microstructure.    Many other approaches try to evade the consequences  of this theorem but ultimately fall foul of it,  ultimately  (and usually implicitly)  violating  (ii) or (iii) in a vain attempt to preserve (i).

\subsection{You can't have your cake and eat it too}
\label{ss:Cake}

The original, core hope of many people  was that small quantum-gravity corrections could save the day.  That is, the growing entanglement in Hawking's leading-order computation would be removed through small sub-leading corrections involving the very large number of  emitted pairs.  For example, Hawking's 2004 retraction of his claim of information loss \cite{Hawking:2016msc} was based such small corrections, caused by a sub-leading saddle point in the Euclidean action.  This idea was based on earlier suggestions in \cite{Maldacena:2001kr}. 

This is precisely the option that was ruled out in the original small corrections theorem  \cite{Mathur:2009hf}: {\it no} small correction to Hawking's process can remove the monotonically growing entanglement of the emitted radiation with the remaining black hole.

Despite this, the same idea still gets re-expressed in different forms. For example, there are discussions of solving the information problem based on complex and subtle measurements of tails of wave-functions. These measurements take extremely long periods of time and require repeated  tests on multiple copies of the black hole.  However, when such arguments are examined more closely \cite{Guo:2021blh}, they simply violate the small corrections theorem, or embrace the giving up of (ii) or (iii).

More recently, it has been suggested that there might be some kind of effective smooth horizon geometry, but that black-hole information can still be recovered \cite{Almheiri:2020cfm}. This idea has a  variety of forms, but it starts with the postulate that unknown quantum gravity effects alter the  structure of the black hole, so that as seen from the outside it appears to behave like a normal body with $A/4G_N$ degrees of freedom, and has a unitary time evolution.  To restore the smooth horizon, it is then supposed that one can find a small ``code subspace'' within the quantum gravitational degrees of freedom in the region of the black hole. The evolution of this code subspace follows the semi-classical gravitational dynamics expected around the horizon on the basis of effective field theory, and, in particular,   creates an effective smooth horizon.

The first postulate means that the information comes out, and there is no information puzzle, despite the presence of the effective horizon created by the code subspace. However, the small corrections theorem has a simple extension to an ``effective small corrections theorem'' \cite{Guo:2021blh}  that shows that this is not possible: one has simply transferred to problem to the code subspace.    The core idea of the improved theorem is that semi-classical effective field theory necessarily generates entangled pairs $b_\eff, c_\eff$ of quanta of the effective fields, and soon after they are created, the radiated quanta $b_\eff$ have propagated far enough from the black hole that they are effectively decoupled from the black hole's internal dynamics, while being maximally entangled with the internal state of the black hole (up to small corrections).  Thus the entanglement of the quanta at infinity keeps rising monotonically, even though the $b_\eff, c_\eff$ were only effective constructs made from the exact degrees of freedom.  

Since this approach has a smooth horizon and posits unitary evolution, it follows that it can only solve the information problem by violating locality through some non-local interaction between the black-hole interior and distant radiation.   Any such modification appears to be highly problematic~\cite{Guo:2021blh}.  The variety of ``island'' scenarios, where the black hole interior is non-locally transferred to the distant radiation field in the course of its evolution, requires such  non-local physics.

In an attempt to avoid such non-local interactions, there have also been suggestions that one must somehow identify the Hawking radiation states with interior states of the black hole \cite{Almheiri:2020cfm}.    The local description of the Hawking process defines the Hilbert space of states of the radiation, and any subsequent identifications amount to acting with a projection operator on that Hilbert space.  Since that projection has a non-trivial kernel, such evolution is {\it ipso facto} non-unitary.  There are also suggestions that one might re-define the Hilbert space inner-product but these re-definitions also violate unitarity.

\subsection{Re-defining the information problem}
\label{ss:Redefine}

One of the more intriguing ideas to resolve the information paradox employs the generalized entropy formula
\be
\label{genent}
S_{\rm gen}(X) = \frac{A(X)}{4G_N} + S_{\rm semi-cl}\big(\Sigma_X\big) ~,
\ee
where $X$ is a {\it quantum extremal surface} (QES) \cite{Engelhardt:2014gca} whose area is $A(X)$, and $S_{\rm semi-cl}$ is the semi-classical entropy of effective quantum fields outside $X$ on the hypersurface $\Sigma_X$ containing $X$.  The particular configuration of $\Sigma_X$ and $X$ are determined by extremizing this expression in a particular way.
It is then argued that the QES contribution dominates the evolution of the entropy during the late stages of evaporation, and it is this property that results in the decrease in entropy of the radiation and resolves the paradox \cite{Almheiri:2020cfm}. 

There are several issues here.  At the most basic level, in order to have the area term in~\eqref{genent} dominate, one has to eliminate the ever-growing entanglement entropy of the Hawking radiation contribution to the second term.
This is achieved by declaring what is ultimately part of the interior of the black hole to belong to $\Sigma_X$, thereby purifying the state of the Hawking radiation.  Such a construction is violently non-local (or else involves non-unitary projections): once the radiation is far from the black hole, it cannot be purified by local effects.  

We have  known since Hawking's original computations that the states needed to purify Hawking radiation exist on space-like slices inside the black hole, and so if one can access those states then one can solve the information problem. By re-defining the radiation entropy to include such internal states, one has re-defined the problem out of existence.

The fact that the QES extremizes a `generalized entropy' of a combined system of radiation and black hole, has led to some interesting Euclidean saddle points, and some suggestion about the Lorentzian dynamics of such systems. However, the generalized entropy that is being extremized in such computations is not the {\it locally defined entropy} of the radiation far from the black hole.  Thus one must ask for the physical implication of finding the QES surface obtained in this manner. It is argued that the value of the generalized entropy follows a Page curve. But this generalized entropy is not the actual entropy of the radiation that appears in the information paradox. Thus one has obtained a Page curve for a quantity that has no relevance to the dynamics of the black hole, and says nothing about the actual entropy whose monotonic rise gives the information paradox. 

Ultimately, the core of the information problem lies in the specific {\it mechanism} by which the radiation state purifies over the course of the black-hole evolution and the QES tells us nothing about this.

\subsection{Incorrect extrapolations of AdS/CFT}
\label{ss:IncorrectAdS}

While AdS/CFT provides remarkable new insights into strongly coupled quantum field theories and black-hole microstructure, there are occasional incorrect extrapolations of these ideas that suggest that unitary, local evolution and horizons are compatible.  They are not!

\subsubsection{Appropriate use}
\label{ss:GoodAdS}

Consider a collection of D3-branes, in their ground state $|0\rangle$.     Suppose a graviton falls onto these branes and is absorbed.  The graviton, and its energy is converted into a collection of open strings (gluons) on the D3 branes.  As time passes, this excitation spreads on the surface of the D3 branes, the evolution being described by strongly-coupled super-Yang-Mills theory. This evolution is very complicated, but it does not involve gravity and is unitary because Yang-Mills theory is unitary.

Now consider this from the perspective of the bulk gravity dual.  The metric produced by the D3 branes is that of  $AdS_5 \times S^5$. The graviton incident on the D3 branes just passes smoothly into the interior if $AdS$.  As the open strings spread on the D-branes  in the SYM description, the graviton in the gravity  description just moves deeper into $AdS$ \cite{Maldacena:1997re}.  

We have chosen a particularly simple example, but this is the essence of holographic field theory, or the AdS/CFT duality.  The dynamics of a gravitational process can be captured completely by a dual boundary theory, which is manifestly unitary (and does not involve gravity).   There is an enormous body of evidence saying that this duality is correct. 

In the fuzzball paradigm, this duality covers black holes in a natural way: there are states in the boundary field theory that can be thought of as dual to microstates of a black hole, and, in the gravity description, these states are just fuzzballs, with no horizon. Thus there is nothing conceptually different between non-black-hole states and black-hole states: any state in the boundary field theory is described in gravity as some excitation of string theory without any horizon.

Indeed,  in the fuzzball paradigm, the black hole is a very messy quantum state and a graviton falling into this object would evolve in a very complicated way as it reaches the horizon radius.  The dynamics of this process can be captured by either the holographic field theory at the boundary (which is unitary, and does not involve gravity), or through the gravitational dynamics of the fuzzball state.  

\subsubsection{Inappropriate use: horizons and information recovery}
\label{ss:BadAdS}

The success of holographic field theory might suggest a different conjecture for black holes:   There could be another description of the gravitational dynamics in which the infalling graviton just passes smoothly into the interior of the horizon, similar to the infall into AdS described above.  This suggests there could be unitarity evolution in the boundary theory and  an effective semi-classical horizon, with semi-classical evolution in some appropriate effective variables in the bulk dynamics.  

However, this is precisely what is excluded by the {\it effective small corrections theorem}.  So where is the error? 

In the correct use of holographic field theory, the D3 branes were in a unique  state $|0\rangle$, and dynamics of the infalling graviton  can be mapped, via the duality,  to the dynamics of perturbations around the unique gravitational state given by the $AdS$ geometry. But in the ``inappropriate use'',  the black hole does not have a unique state: it has $\it exp[S_{Bek}]$ states, and such a vast number of states  {\it cannot} be replaced by a {\it single} geometry without a huge loss of information. The single black-hole geometry has no knowledge of the  details of the microstructure, and any computation done with this geometry will necessarily manifest information loss.

Black holes, do, of course have a well-understood holographic dual, and do have an appropriate use:  the exterior geometry of the black hole provides a way to describe the gravitational dual of a thermal field theory.   The error of ``inappropriate use'' lies in replacing a single state by the semi-classical geometry, and then hoping to extract the detailed microstructure by using the interior region of this geometry.

This erroneous extrapolation of AdS/CFT is sometimes compounded by another error: Applying the duality to the semi-classical geometry of the eternal black hole.  This geometry has {\it two} boundaries, so the dual CFT is described by two theories that are non-interacting, though they can be in an entangled state. Since the gravity solution connects the two CFTs by a `wormhole' (the Einstein-Rosen bridge), it appeared to people that entanglement can generate physical connections through which information may pass.   It must be emphasized that entanglement simply means taking specific linear combinations of states in a Hilbert space, and is nothing more than a change of basis.  Wormholes impute far more:  a long-range, non-local interaction Hamiltonian (as, for example, in \cite{Maldacena:2018lmt}).

 It can be shown, however, that applying holographic duality to semi-classical geometry of the eternal black hole leads to contradictions with unitarity of the eternal black hole \cite{Mathur:2014dia}.  Moreover, in the fuzzball paradigm,   the error in the wormhole argument is clear.  Recall that  the single sided black-hole geometry is invalidated by tunneling into fuzzballs. Similarly, the two-sided eternal black hole geometry is also invalidated by tunneling into pairs of fuzzball states, yielding  disconnected (but entangled) fuzzballs\footnote{A similar picture appears to emerge from the analysis of \cite{Jafferis:2021ywg}.}.
 If we assume that the single-sided black hole exists, even in an approximation, then the Hawking argument implies loss of unitarity.  Similarly, any conclusions using the eternal black hole geometry will also lead to inconsistent physics. Thus it is not surprising that all pictures derived from the wormhole idea turn out to be inconsistent.
 
\subsubsection{Inappropriate use: motivating non-local gravitational dynamics}
\label{ss:Nonlocality}

Holographic duality provides a map between states and dynamics of a boundary gauge theory and the states and dynamics of a bulk gravitationally-coupled theory.   Of necessity, this map is highly non-local: It maps local excitations deep in the interior of the spacetime geometry to complicated strongly correlated, highly distributed excitations of the gauge theory.  Thus, it might be argued that since AdS/CFT involves non-locality, we must embrace non-locality within our theories of Nature.

This claim misses the point entirely.  In holographic field theories, the dynamics of the boundary theory is entirely local {\it and} the dynamics of the bulk gravitational theory is also entirely local.  The only non-locality inherent in AdS/CFT is in the mapping between bulk and boundary.   At no point does holographic field theory, or AdS/CFT, tell us that we must give up locality in the field theories in either the bulk or on the boundary.   In particular, there is no holographic reason that tells us that the gravitational dynamics in the bulk is in any way non-local.%
\footnote{Of course in string theory there is an intrinsic non-locality at the string scale, but this is not so much non-locality as much as having objects whose wavefunction has a minimum size.  The Virasoro constraints pull back the spacetime light-cone structure to the string worldsheet, and ensure that there is no a-causal propagation.  Fuzzballs are similarly not so much non-local objects; rather, they are again objects with a minimum size (the Schwarzschild radius) over which their quantum wavefunction is coherently distributed.}

\section{The firewall muddle}
\label{App:B}

The firewall argument of 2012 \cite{Almheiri:2012rt} created an enormous amount of confusion. The argument said: If we require the entanglement of Hawking radiation to start going down at some point in the evaporation process, then  infalling observers will not find the horizon to be a vacuum: they will see a firewall. 

The first point of confusion was: how is this different from Hawking's original paradox? It isn't; and to see why it is worth recalling a little history. 

\subsection{QFT in curved spacetime }
\label{ss:Boulware}

Hawking argued that the collapse process that leads to black-hole formation leaves behind a near-vacuum (Unruh) state at the horizon, where the fields are to be quantized  in inertial coordinates.   The choice of coordinate affects the expectation value of the stress tensor (see~\cite{Birrell:1982ix} for a review).%
\footnote{For instance, one can regularize the stress tensor by point-splitting the fields.  The resulting OPE singularities are determined by the UV asymptotics of the continuum quantum field theory.  Subtracting the singular part in different coordinates leads to different renormalization prescriptions.} 
Doing the subtraction in inertial coordinates leaves a sensible, finite result for the stress tensor expectation value at the horizon; doing the subtraction in Schwarzschild coordinates, which are singular at the horizon, leads to a singular stress tensor expectation value there.  

These different choices of subtraction define different vacuum states of the field~-- the Unruh vacuum for inertial coordinates, and the {\it Boulware vacuum} for Schwarzschild coordinates.

Studies of flat space in Rindler coordinates revealed much of the physics of how these states differ (see for example \cite{Unruh:1983ms,Birrell:1982ix}).   From the perspective of uniformly accelerated (Rindler) observers, the (smooth) Minkowski vacuum is a very complicated combination of excited Rindler states:
\begin{equation}
\label{Rindler1}
\big|0\big\rangle_{\!\it Minkowski} ~=~  \prod_j \Bigg[ \frac{1}{(1- e^{-\pi n_j \omega/a})^{1/2}} \, \sum_{n_j} \, \big| n_j, L  \big\rangle \otimes \big| n_j, R  \big\rangle  \Bigg] 
\end{equation}
where $ \big| n_j, L  \big\rangle $ and $\big| n_j, R  \big\rangle$ denote  the state with $n_j$ particles of frequency $\omega_j$ in the left and right Rindler wedges.  The smoothness of the Minkowski vacuum across the Rindler horizon critically depends on the precise, and very delicate, correlation of excited states in both Rindler wedges.  If one disturbs this correlation, even by a small amount, the resulting state is ``pathological'' at the Rindler horizon because the perturbation persists to arbitrarily short distances/high frequencies.%
\footnote{The different coordinate systems lead in the operator formalism to different notions of normal ordering, and therefore different notions of vacuum state, and normal ordering prescriptions for the stress tensor.}  

This analysis provides an excellent model for the vacuum structure of a black hole near the horizon, with the infalling vacuum playing the role of the Minkowski vacuum (smooth across the horizon), and the Boulware vacuum playing the role of the Rindler vacuum.  The infalling vacuum is thus a very delicate correlation of excited states from the perspective of Schwarzschild time, and disturbing this precise correlation of states results in a singular state at the horizon, and the most extreme choice is the mother of all firewalls,  the Boulware vacuum.  

Put simply, disturbing the infalling vacuum at the horizon by tinkering with the Schwarzschild modes is utterly pathological because it puts some disturbance at the horizon, and since the blue-shift is infinite at the horizon this  disturbance necessarily involves unbounded frequencies and infinite energies: a firewall.   

Thus, at its most basic level,  the recent firewall story is simply trying to re-litigate a forty-year-old discussion of whether one should choose the infalling vacuum, the Boulware vacuum, or some variant thereof.   Nearly 50 years ago, Hawking settled the ``firewall debate'' by making a  compelling case that the only sensible choice, at least in GR, is to choose a vacuum state that is smooth at the horizon.

This old story is being revisited in the form of a firewall because the entanglement of the Hawking radiation will only come down if the horizon is not the infalling vacuum. The `firewall paradox' is simply a rephrasing of the original Hawking paradox.

\subsection{The firewall}
\label{ss:firewall}

In 2009, the small corrections theorem  \cite{Mathur:2009hf}  sharpened the Hawking paradox, and reduced the issue to the choices summarized at the beginning of Appendix \ref{App:A}.  This result was enlisted in \cite{Almheiri:2012rt} in support of an argument that the horizon region could not be smooth, and, in particular it must be some kind of firewall.   

The firewall arose because the analysis was limited to semi-classical GR and the use of  effective field theory at the horizon.  The basic idea was that, if the information was to escape the black hole, then ``late radiation'' must be entangled with ``early radiation'' and so cannot be entangled with interior states of the black hole.  This loss of entanglement across the horizon destroys the delicate correlations inherent in a vacuum at the horizon and, for the reasons discussed above, one concludes that, at least in an ``old black hole'', there must be a firewall at the horizon.  

This analysis provides an illustration of the broader issues raised by Hawking, and further refined in  \cite{Mathur:2009hf}. As we will discuss below, the need for a firewall reflects the paucity of options provided by an analysis in GR alone.  The real answer to this problem comes from string theory, and is the fuzzball.  

The other core message in  \cite{Almheiri:2012rt} was to re-visit another, earlier confusion known as Black Hole Complementarity \cite{tHooft:1984kcu,Susskind:2005js, Susskind:1993if}.  The idea of Black Hole Complementarity was to posit that, when described in one set of observables, the black-hole information would be returned, or reflected, from the black-hole horizon via some unspecified mechanism, and yet, in some complementary set of quantum observables, one would still see a smooth horizon and an infalling observer would feel a continued free fall beyond it.  The problem with this idea is that, in this second set of observables, the small corrections theorem still applies and so, once again, the information cannot be returned without sacrificing one of the three things listed at the beginning of Appendix \ref{App:A}.  As was noted in the exposition given in \cite{Almheiri:2012rt}, `complementarity is not enough.'   The firewall argument simply pointed out that Hawking's problem remains a problem even with the complementarity conjecture.

It is  interesting to note that  the ideas of effective smooth horizon geometry and small ``code subspaces'' described in Appendix \ref{ss:Cake}, have much in common with the ideas of Black-Hole Complementarity, and both must fail for the same reason, either at the hand of the small corrections theorem, or  the more general, effective small corrections theorem.

There is, however, much to sympathize with in the ideas of firewalls and black-hole complementarity, and we believe that one or the other might be realized, in some form, within the framework of fuzzballs.

\subsection{Fuzzball and horizon-scale physics}
\label{ss:HorizonFuzz}

As much of this paper describes, fuzzballs render firewalls and black-hole complementarity entirely moot because fuzzballs have no horizon, and the radiation does not emerge from pair creation from the vacuum.  It is, however interesting to investigate how and why fuzzballs succeed where other approaches have fallen short.  As we will describe, the answer lies in the ``fuzzball principle:"  Horizons and singularities only appear if one tries to describe gravity using a theory that has too few degrees of freedom to resolve the physics.   We would add that some of the problems in other approaches also come from failure to incorporate gravitational dynamics and, in particular, correctly address gravitational back-reaction.  
 
As described in Section \ref{ss:FuzzFormation}, the ``elephant in the room,''  is the vast density of states present in a black hole.  This has a huge influence on the effective horizon-scale physics.  GR is incapable of capturing even the crudest approximation to the dynamics of such a system: it can only describe meagre averages, and so gives us horizons, singularities, firewalls and  the information problem.

String theory has shown that it can capture the microstructure of black holes, and the massless limit, supergravity, can capture some of the large-scale effective dynamics of horizon-scale physics. Microstate geometries can support horizon-scale structure, but GR cannot.  Therefore any matter representing a firewall will be instantly swept into the interior of a black hole in GR.  Conversely, the only way to make semi-classical sense of a firewall is as `hot fuzz', supported by a microstate geometry.

Intriguingly enough, there is also an avatar of  black-hole complementarity that is resurrected in {\it fuzzball complementarity}. While still an open conjecture, fuzzball complementarity illustrates many aspects of how the fuzzball paradigm goes well beyond the limits of GR.

\subsubsection{Fuzzball complementarity}
\label{ss:FuzzComp}

Fuzzball complementarity  proposes that, while  emitted quanta  with energies comparable to the temperature,  $E\sim T$, emerge from the fuzzball in a state that depends on the details of the fuzzball, some features of the dynamics of modes with  $E\gg T$ could be described {\it approximately} by using the traditional geometry of the black hole \cite{Mathur:2012zp,Mathur:2014dia}.   

At first the firewall argument claimed that in addition to ruling out black-hole complementarity, it had also ruled out fuzzball complementarity. But this claim was incorrect \cite{Mathur:2013gua}, and a lot can be learnt by understanding where the argument went wrong.   To see the error, we have to realize that the firewall argument  tried to make  Hawking's conclusion more pointed by adding an extra assumption, and, when it comes to fuzzballs, it is this assumption that proves to be the argument's undoing. 

As we have noted above, Hawking's argument can be restated as:

\medskip

{\it If the entanglement has to come down, then the horizon cannot be the vacuum.}

\medskip
\noindent However, as far as this argument goes, the departure from the vacuum can happen at any length scale $l_p\lesssim \lambda \lesssim R_s$. The firewall argument added the extra assumption:

\medskip

{\it  It is assumed that effective field theory (EFT) is valid for $r>R_s+l_p$.}

\medskip
\noindent
This assumption confines the required changes of the vacuum state to Planck-scale modes around  the horizon. Since these modes have high energy, it is then stated that an infalling observer interacting with these modes will feel a firewall. The first flaw in this argument \cite{Mathur:2013gua} comes from the failure to address gravitational back-reaction. 

Consider an infalling quantum with $E\gg  T$. The back-reaction of the infalling quantum creates a {\it new} horizon that is {\it outside} the old horizon $R_s$, at some radius $R'=R_s+d$ with $d\gg l_p$. Now we note two facts:

\begin{enumerate}[(a)]
\item
The temperature at distance $d$ from the horizon is $T_d\sim 1/d$. Thus while the infalling quantum had an interaction probability of order unity with the thermal quanta  at $R_s+l_p$, this probability of interaction is small at the new horizon $R'$ for $E\gg T$. In other words, the local temperature is too low at the new horizon to give `firewall' behavior.
\item
Effective Field Theory must break down at the {\it new} horizon $R'$. This is because light cones turn `inwards' at $R'$, and if semi-classical physics remained valid at $R'$, then no novel physics at $r<R'$ can alter the physics at $R'$. Continuous pair creation will then necessarily happen at $R'$ and will invalidate the requirement that the entanglement starts decreasing at some point. This breakdown brings in new degrees of freedom that are not included in semi-classical dynamics, and it is the dynamics of these  that can give `fuzzball complementarity'. 
\end{enumerate}
To see how (b) bypasses the firewall argument, note that  the argument was based on Hawking's pair creation
\be
|\psi\rangle_{pair}={1\over \sqrt{2}} \Big ( |0\rangle_b|0\rangle_c+|1\rangle_b|1\rangle_c \Big )
\label{appbone}
\ee
If the entanglement is to go down, then the radiated quantum $b$ has to be entangled with the  earlier radiated quantum $\{ b'\}$, and so cannot be entangled with its partner $c$ in the form (\ref{appbone}) required for a vacuum at the horizon. But now consider the infall of a quantum with $E\gg T$. Before the infall, the hole with mass $M$ had entropy $S_{\it \it Bek}(M)$. After the infall, this entropy will be 
\be
S_{\it Bek}(M+E)\approx S_{\it Bek}(M)+{dS_{\it Bek}(M)\over dM} E = S_{\it Bek}(M) +{E\over T}
\ee
Thus the number of states after the infall are far larger than the number of states before the infall
\be
{N_{\it after}\over N_{\it before}}=\frac{\exp[S_{\it Bek}(M+E)]}{ \exp[S_{\it Bek}(M)]}= e^{E\over T} \gg 1
\ee
Note that there are a vast number of new states:  $e^{E/ T}$ {\it multiplies} $N_{\it before}$. These newly created states {\it are not entangled with anything}, and it is the dynamics of these new states that allows the conjecture of fuzzball complementarity.  The question becomes: How does the infalling observer interact with this vast collection of degrees of freedom lying outside of effective field theory. If they impart only small kicks to infalling observer for some period of time, then the experience might be like that of Black Hole Complementarity; if they impart large momentum transfers at a rapid rate, then the experience is akin to a firewall.

While one cannot establish the fuzzball complementarity conjecture without knowing more about the dynamics of fuzzballs, a bit model exhibiting fuzzball complementarity was made in \cite{Mathur:2015nra}. In this model the Page curve is that of a normal body, while an infalling quantum observer with $E\gg T$  sees `no drama' at the horizon. For our discussion in this review, fuzzball complementarity illustrates, and rests upon, two crucial features that are often neglected in other analyses: (i) gravitational back-reaction and (ii) the dynamics of the vast number of states present in the black hole, or fuzzball.


\begin{adjustwidth}{-1mm}{-1mm} 

\bibliographystyle{utphys}

\bibliography{microstates}       

\end{adjustwidth}

\end{document}